\documentclass[12pt]{iopart}

\usepackage{graphicx} 
\usepackage{epsfig}
\begin{document}

\title[The $N=28$ shell closure: a test bench for nuclear forces]
{Evolution of the $N=28$ shell closure: a test bench for nuclear forces}

\author{O. Sorlin$^1$ and M.-G. Porquet$^2$}

\address{$^1$Ganil, CEA/DSM-CNRS/IN2P3, B.P. 55027, F-14046 Caen Cedex 5,
France}
\address{$^2$CSNSM, CNRS/IN2P3 - Universit\'e Paris-Sud, F-91405 Orsay,
France}
\ead{sorlin@ganil.fr,porquet@csnsm.in2p3.fr}
\begin{abstract}
The evolution of the $N=28$ shell closure is investigated far from
stability. Using the latest results obtained from various experimental
techniques, we discuss the main properties of the $N=28$ isotones, as
well as those of the  $N=27$ and $N=29$ isotones. Experimental 
results are confronted to various theoretical predictions.
These studies pinpoint the effects of several terms of the nucleon-nucleon
interaction, such as the central, the spin-orbit, the tensor and the
three-body force components, to account for the modification of the $N=28$ 
shell gap and spin-orbit splittings. 
Analogies between the  evolution of the $N=28$ shell closure and other 
magic numbers originating from the spin-orbit interaction are proposed 
($N=14,50, 82$ and $90$). More generally, questions related to the 
evolution of nuclear forces towards the drip-line, in bubble nuclei, 
and for nuclei involved in the r-process nucleosynthesis are proposed 
and discussed. 

\end{abstract}

\pacs{21.10.-k, 21.30.-x, 21.60.-n, 27.30.+t, 27.40.+z}
%\submitto{\PS}
%\maketitle

\section{Introduction\label{intro}}

Guided by the existence of a spin-orbit coupling in atomic physics, 
M.~Goeppert Mayer~\cite{Goep49} and O.~Haxel \etal~\cite{Haxe49}
independently postulated in 1949 the existence of a strong spin-orbit 
(SO) force in the one-body nuclear potential to reproduce the so-called 
`magic numbers' above $20$. Their SO interaction has to be attractive  
for nucleons having their angular momentum aligned with respect to 
their spin (denoted
as $\ell _\uparrow$) and repulsive in case of anti-alignment 
($\ell _\downarrow$). Its strength was constrained to reproduce
the size of the shell gaps at 
$28$, $50$, $82$, and $126$ derived from experimental data
obtained in the early days of nuclear structure studies.

During the last two decades, major experimental and theoretical 
breakthroughs have strengthened that, though this simplified 
mean field picture  of the SO interaction is elegant and simple, 
the observed SO splitting comes from the combination of complex 
effects due to the nuclear force. 
In particular, the size of the $N=28$ shell gap is governed by
the spin-orbit, the tensor, 
the three-body force components, and possibly the change of nuclear forces 
at the drip-line due to interactions with the continuum. For instance,  
the $N=28$ shell gap grows by about 2.7~MeV between $^{41}$Ca 
and $^{49}$Ca to reach a value of 4.8~MeV. This effect has 
recently been mainly inferred to three-body forces.  Then, starting 
from the doubly-magic $^{48}_{20}$Ca$^{}_{28}$ nucleus and removing
protons, a progressive onset of deformation is found to occur: 
$^{46}_{18}$Ar 
is a moderately collective vibrator, $^{44}_{16}$S exhibits 
a prolate-spherical shape coexistence,  the  $^{42}_{14}$Si nucleus is 
a well oblate nucleus, and the newly discovered $^{40}_{12}$Mg 
is likely to be a prolate  nucleus. Such a progressive change 
of structure is triggered by the combination of three effects: 
the reduction of the spherical $N=28$ shell gap, the probable reduction of the $Z=14$ sub-shell gap
 \emph{and} the increase of quadrupole excitations across these gap, i.e. on the one side between 
the neutron $1f_{7/2}$ and $2p_{3/2}$ orbits and on the other side between the proton $1d_{5/2}$ and $2s_{1/2}$ orbits which are 
separated by two units of angular momenta (2$\hbar$). 

This smooth but continuous change of nuclear shapes at the 
semi-magic $N=28$ nuclei is quite unique in the chart of 
nuclides. Therefore from this study we can learn how a spherical 
rigid system could evolve towards a very collective system 
when a few nucleons are removed. There, 
proton-neutron forces act between nucleons in orbits which are 
different from the ones occupied in the valley of stability, 
hereby involving new facets of the nuclear forces which were 
not yet tested. 

The magic number $28$ is in fact the third number of nucleons 
which is  created by the SO splitting\footnote{Note that from now on, we 
will rather explicitly mention SO \emph{splitting} rather than SO 
\emph{interaction} as the shell gaps are created by several 
components of the nuclear force, and not exclusively by the 
spin-orbit force.}, the first two SO gaps being $6$ and $14$. 
Similar components of the nuclear forces are expected to play a 
role in the gap values of all the SO magic numbers, leading 
to possibly similar consequences throughout the 
chart of nuclides. The magic number $28$ is a target choice to 
study nuclear forces as it involves nuclei having an intermediate 
mass and size, with orbits relatively well separated from the 
neighboring ones. This feature is essential to distinguish the 
effects of nuclear forces between nucleons occupying orbits having 
different properties (such as the number of nodes, the angular 
momentum and the spin orientation). Conversely as the size of 
nuclei grows, the in-medium nuclear force is weaker and modification 
of shell structures occur only when many nucleons are involved. 
Added to this, the nuclear orbits are more compressed in energy 
and their occupancy is diluted by nucleon-pair scatterings. It follows 
that the action of specific nuclear forces between nucleons in 
well established orbits is much more difficult to disentangle in heavy
than in light systems. At the other extreme, if the system is 
composed of a small number of nucleons,  the change of only few nucleons leads to very rapid 
and drastic structural changes and the very concept of the mean-field approach becomes inappropriate. 
To quote one example of quick change of a SO magic gap, the $N=14$ gap collapses already after 
the removal of two protons, from $^{22}$O to $^{20}$C. 

On the experimental point of view, an incredible wealth of 
information has been obtained these last two decades by studying the 
evolution of nuclear structure south to the doubly-magic 
$^{48}_{20}$Ca$^{}_{28}$ nucleus. Different experimental techniques 
are used 
worldwide, such as $\beta$ or isomeric-decay studies, atomic 
mass measurements, Coulomb excitation, transfer reaction, or 
in-beam $\gamma$-ray spectroscopy. These studies became feasible 
since increasing beam intensities of $^{48}$Ca and detectors capabilities become available. 
With 8 neutrons more than the most abundant $^{40}$Ca 
isotope, the  $^{48}$Ca isotope represents a fraction about 0.2\% 
of the total Ca isotopes on earth. However it is 
more than 50 times more abundant than the $(N-2)$ nucleus 
$^{46}$Ca on earth, a feature which is not so common in the 
chart of nuclides. This is ascribed to its double magicity 
which makes it  produced in significant amount and poorly 
destroyed in specific stellar explosive environments. In some 
sense, nature has offered with  the neutron-rich and doubly magic 
$^{48}$Ca isotope a rather 
unique tool to study the behavior of nuclear forces far from 
stability.

%%%%%%%%%%%%%%%%%%%%%%%%
\begin{figure}
\centering \epsfig{height=11cm, file=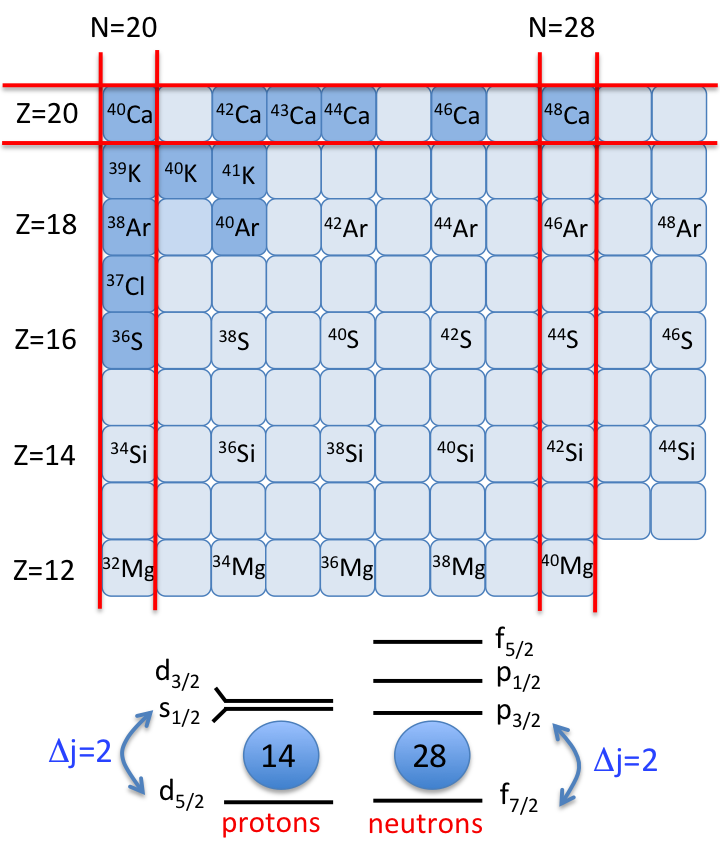} 
\caption{Part of the chart of the nuclides which will be considered 
in the present work. Stable nuclei are displayed with a dark blue color. 
In the bottom part, the orbits of the proton and neutron valence spaces 
under study are shown.}
\label{Fig1}
\end{figure} 
%%%%%%%%%%%%%%%%%%%%%%%%
We shall explore in the next Sections the nuclear structure evolution
 of the nuclei presented in Figure~\ref{Fig1}. In Section~\ref{view},  the 
general properties of atomic masses, first excited states and 
reduced transition probabilities will be presented. In
Section~\ref{proton_espe}, 
we shall discuss upon the  evolution of proton orbits when 
moving from $N=20$ to $N=28$ by studying the odd-Z isotopic 
chains of $_{19}$K, $_{17}$Cl and $_{15}$P. 
Sections~\ref{sectN27} and \ref{neutron_espe}
show the evolution of the $N=28$ gap from the behaviors of 
the $N=27$ and $N=29$ isotones. Some 
experimental and theoretical challenges related to the nuclear 
forces responsible for the SO magic numbers will  be proposed in the last Section.

Sections 2 to 5 of the present paper update our previous work on 
the evolution of the $N=28$ shell closure~\cite{nous}.

\section{View of the shell-structure evolution from atomic masses, 
$E(2^+_1,0^+_2,4^+_1)$ and $B(E2; 0^+_1 \rightarrow 2^+_1)$ 
values}\label{view}

\subsection{Atomic masses}\label{masses}
\begin{figure}
\centering \epsfig{height=8cm, file=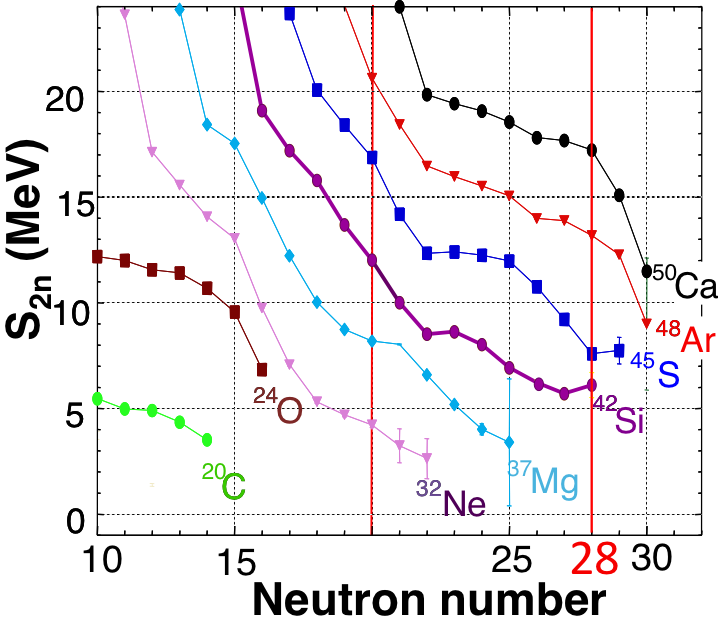} 
\caption{Evolution of the two-neutron separation energy 
$S_{2n}$ in the even-Z isotopic chains 
(experimental atomic masses from Refs.~\cite{Jura07,AMDC12}). The 
sharp drop in $S_{2n}$ observed in the Ca isotopic chain 
after $N=28$ progressively disappears far from stability.}
\label{S_2n}
\end{figure}

The two-neutron separation energy $S_{2n}(A)$ can be obtained 
in a given isotopic chain from the atomic-mass differences 
between nuclei having $A$ and $A-2$ nucleons. The trend of $S_{2n}(A)$ is in 
general rather smooth, except after having passed a major 
shell gap. There, when adding 2 more neutrons,  
the $S_{2n}(A)$ value drops significantly. This is an indication 
that a shell closure is efficient and that a spherical shell gap 
is existing. Such a sharp drop in $S_{2n}(A)$ is observed in the 
Ca and Ar isotopic chains (see Figure~\ref{S_2n}).
%%%%%%%%%%%%%%%%%%%%%%%%%%%%%%%%%%%%%%%%
%%%%%%%%%%%%%%%%%%%%%%%%%%%%%%%%%%%%%%%%
On the other 
hand, the S and Si isotopic chains behave differently.
An \emph{increase} in $S_{2n}(A)$ is observed in the S chain when crossing $N=28$  
and in the Si chain, just before crossing $N=28$ . This 
indicates a clear deviation to sphericity at $N=28$ in these isotopic chains. 
This increase of $S_{2n}(A)$ around $N=28$  likely arises from the increase of binding energy due to correlations, 
which goes in concert with the onset of deformation. These hypotheses  need to be confirmed with other nuclear 
properties such as the first $2^+_1$ energy  and the reduced transition 
probability $B(E2;0^+_1 \rightarrow 2^+_1)$. 

\subsection{Evolution of $E(2^+_1)$ and 
$B(E2;0^+_1 \rightarrow 2^+_1)$ values in the $N=20-28$ nuclei}\label{evol}

\begin{figure}[h!]
\begin{minipage}{8cm}
\begin{center}
\includegraphics[width=7cm]{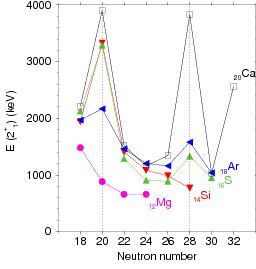}
\end{center}
\end{minipage}\hfill
\begin{minipage}{8cm}
\includegraphics[width=6.7cm]{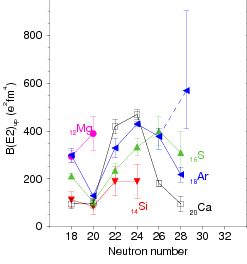}
\end{minipage}
\begin{center}
%\begin{minipage}{16.5cm}
\caption {Experimental $E(2^+_1)$ energies (left) and $B(E2; 0^+_1
\rightarrow 2^+_1)$ values, also noted $B(E2)_{up}$ or 
$B(E2)$$\uparrow$, (right) in the $_{12}$Mg to $_{20}$Ca isotopic
chains as a function of the neutron number $N$. Data are taken
from the compilation of Ref.~\cite{Rama01}, except for
$^{38}$Si~\cite{Lian03}, $^{40}$Si~\cite{Camp06},
$^{42}$Si~\cite{Bast06}, $^{40}$S~\cite{Wing01}, $^{46}$S~\cite{Gade09},
$^{44}$Ar~\cite{Ziel09}, $^{46}$Ar~\cite{Gade03,Meng10},
and $^{48}$Ar~\cite{Bhat08}. 
}
\label{E2BE2N2028}
%\end{minipage}
\end{center}
\end{figure}

Most generally, magic (spherical) 
nuclei are characterized by high $E(2^+_1)$ and weak 
$B(E2;0^+_1 \rightarrow 2^+_1)$ values. 
Any deviation from this trend reveals 
a change in nuclear structure.  The systematics of the $E(2^+_1)$ and B(E2) values from $^{40}$Ca to
$^{48}$Ca (see  Figure~\ref{E2BE2N2028}) is often used to model 
multi-particle configurations of valence
nucleons in one particular orbit, which is the $\nu f_{7/2}$ orbit
in the present case. In this framework, the 2$^+_1$ energies 
are maximum at the beginning  of the shell ($N=20$) where there is no valence neutron,  and the end of the shell ($N=28$) where neutron excitations can no longer occur in the same shell. The  2$^+_1$ energies are almost constant in between. In parallel 
the B(E2) values follow a bell-shape curve with minima at the two extremes of the shell according to the relation:

%%%%%%%%%%%%%%%%%%%%%%%%%%%%%%%%%%%%%%%%
\begin{equation}\label{BE2}
 B(E2) \propto F(F-1)
\end{equation}
%%%%%%%%%%%%%%%%%%%%%%%%%%%%%%%%%%%%%%%%
where F is the fractional filling of the shell, which amounts here to $F=(N-20)/8$.

A deviation to this curve would be due to the 
breakdown of a spherical shell gap and an onset of 
deformation. Importantly, it could also happen when the configuration of the $2^+_1$ state changes 
from a neutron origin to proton one in a given isotopic chain. In this case
a sudden enhancement in $B(E2)$ value can be observed as protons 
carry a much larger effective charge than neutrons (typically $e_p$=1.5 while $e_n$=0.5). 
Added to the $E(2^+_1)$  and $B(E2)$ values,
the 4$^+_1$ and $0^+_2$ energies provide 
important complementary information on the evolution of the shell 
structure. The ratio $E(4^+_1/2^+_1)$ is often used as an 
indicator of deformation, while the location of the 
$0^+_2$ state and the reduced electric monopole strength 
$\rho (E0;0^+_2 \rightarrow 0^+_1)$ are used to probe shape coexistence. 
%%%%%%%%%%%%%%%%%%%%%%%%%%%%%%%%%%%%%%%%%%%%%%%%%%%%%%%%%%%%%%%%%%%%%%
\vspace{0.5cm}
%%%%%%%%%%%%%%%%%%%%%%%%%%%%%%%%%%%%%%%%%%%%%%%%%%%%%%%%%%%%%%%%%%%%%%%%%

Figure~\ref{E2BE2N2028} shows the evolution of the 2$^+_1$ energies 
and of the reduced transition probabilities $B(E2; 0^+_1 \rightarrow 2^+_1)$ 
as a function 
of the neutron number in the $_{12}$Mg, $_{14}$Si, $_{16}$S, 
$_{18}$Ar and $_{20}$Ca isotopes.
At $N=20$, the $2^+$ energies are high for all isotones, 
except for $^{32}_{20}$Mg, in which a drop in energy 
is observed instead. The B(E2) values are  small and remarkably 
similar for all $N=20$ isotopes, 
except again for $^{32}_{20}$Mg. This picture is consistent with a
strong shell closure $N=20$ which persists from $Z=20$ to $Z=14$, 
and disappears below $Z=14$ (see \cite{Kanungo} for further discussions). 
On the other hand, the behavior of the 
$N=28$ isotones is different.  
The rise in $2^+_1$ energies observed at $Z=16-18$ is 
more modest than at $N=20$. In particular, the 
rise at  $^{44}$S$_{28}$ is much smaller than at $^{36}$S$_{20}$. 
In $^{42}_{14}$Si$_{28}$ a drop is observed,  
at variance with what was observed in the 
$^{34}$Si$_{20}$ isotope. Combining this view of the $2^+$ energy 
trend and the evolution of $S_{2n}(A)$, one can surmise that the 
$N=28$ shell closure is progressively vanishing below Ca. The trend 
in $B(E2)$ nicely confirms this statement, as
discussed in the next Section.

\subsection{Spectroscopy of the even-Z $N=28$ isotones\label{N28trend}}

In this Section, we discuss more precisely some properties of 
the even-Z $N=28$ isotones, from $Z=18$ to $Z=12$.

\subsubsection{{\bf $^{46}$Ar}}
 
As the Ar isotopes have an open proton shell configuration, the 2$^+_1$ state is likely to be \emph{mainly} of proton origin
there.  The 2$^+_1$ energy at the $N=20$ and $N=28$ shell closures is smaller in the Ar isotopes
as it comes naturally from the proton re-coupling inside the $sd$ shells. However, a rise in $2^+$ energy is found,
witnessing the presence of neutron shell closures. It is worth noting 
that two very different values of $B(E2; 0^+_1 \rightarrow 2^+_1)$ values 
have been determined in $^{46}$Ar (Figure~\ref{E2BE2N2028}). 
They arise from two different experimental techniques.
On the one side, two 
experiments were carried at the NSCL facility to study the  
Coulomb excitation of $^{46}$Ar. They led to consistent results, 
with a rather small $B(E2)_{up}$ value of 218(31) e$^2$fm$^4$~\cite{Gade03}. 
On the other side, a fairly large $B(E2)_{up}$ value of  570 
$^{+335}_{-160}$ e$^2$fm$^4$ was  deduced from another experiment 
performed at the Legnaro 
facility by Mengoni \etal\cite{Meng10}. In this work, the 
$^{46}$Ar was produced by means of multi-nucleon transfer reaction. 
The lifetime of the $2^+_1$ state was determined by using the 
differential recoil distance Doppler shift method. In addition, 
a tentative 4$^+_1$ state has been proposed at 3892~keV by 
Dombr\'adi \etal\cite{Domb03}, 
leading to E(4$^+_1$/2$^+_1$)=2.47, a value which lies between 
the limits of vibrator and rotor nuclei. 

Comparison of these two $B(E2)$ values to models is very 
instructive. The recent relativistic Hartree-Bogoliubov model 
based on the DD-PC1 energy density functional~\cite{Li11} 
reproduces very well the low value of Ref.~\cite{Gade03}, while the 
Shell-Model (SM) calculations of Nowacki \etal~\cite{Nowa09} 
agree with the large value of Ref.~\cite{Meng10}. It therefore seems 
that, though only two protons below the doubly-magic $^{48}$Ca nucleus,  
$^{46}$Ar is still a challenging nucleus to modelize. 

Interesting to add is the fact that the B(E2) in $^{47}$Ar~\cite{Wink12} 
is close to the lowest value measured in
$^{46}$Ar, i.e. at variance with the SM predictions, while the $B(E2)$ experimental result 
and SM calculations agree with an enhanced collectivity for $^{48}$Ar~\cite{Wink12}. SM calculations also account for the spectroscopy 
of higher energy states in $^{48}$Ar obtained from 
deep-inelastic transfer reactions, which show that its deformation is likely nonaxially symmetric~\cite{Bhat08}. 
To conclude, a new measurement of the B(E2) in $^{46}$Ar is therefore called for to see if a local discrepancy 
between SM calculations and experimental results persists at $N=28$.

\subsubsection{{\bf $^{44}$S}}

The $B(E2; 0^+_1 \rightarrow 2^+_1)$ and $2^+_1$ energy values 
in $^{44}$S, determined using Coulomb excitation at high energy 
at the NSCL facility by Glasmacher \etal~\cite{Glas97}, point 
to a configuration which is intermediate between a spherical 
and deformed nucleus. Evidence of shape coexistence was found 
at GANIL by determining the decay rates of the isomeric $0^+_2$ 
state at 1365(1) keV to the $2^+_1$  at 1329(1) keV and $0^+_1$ 
ground state through delayed electron and $\gamma$ 
spectroscopy~\cite{Forc10}. 
Comparisons to shell-model calculations point towards 
prolate-spherical shape coexistence. A schematic two-level mixing model 
was used to extract a weak mixing between the two configurations, the
0$^+_1$ state having a deformed configuration (i.e. a two-neutron
particle-hole $2p2h$ excitation across the $N=28$ spherical gap)
and the 0$^+_2$ state, a spherical configuration 
(i.e. a zero-neutron particle-hole excitation $0p0h$).
New excited states in $^{44}$S were found afterwards~\cite{Sant11,Cace12}, which lend 
support to the shape coexistence hypothesis. By using the two-proton knockout 
reaction from $^{46}$Ar at intermediate beam energy at NSCL, four 
new excited
states were observed~\cite{Sant11}. The authors proposed that one of them 
has a strongly deformed configuration, due to the promotion of a
single neutron ($1p1h$) across the $N=28$ gap. It would follow that
three configurations would
coexist in $^{44}$S, corresponding to zero-, one-, and two-neutron
particle-hole excitations. The one-proton  knockout 
reaction from $^{45}$Cl gave access to other excited states and their   
$\gamma$ decays were measured using BaF$_2$ detectors at 
GANIL~\cite{Cace12}. In particular the existence of the 
2$^+_2$ state at 2156(49)~keV is confirmed and is likely the expected
'spherical' 2$^+$ state.  The Hartree-Bogoliubov model 
based on the DD-PC1 energy density functional~\cite{Li11} gives a relatively good 
agreement with experimental results, with the exception of a  too large mixing between the 0$^+_1$ and 0$^+_2$ states.
The Gogny force with the D1S set of parameters also gives a good description 
of $^{44}$S, as described in Refs.~\cite{Peru00,Guzm02}.
Shape coexistence has also been evidenced in the neighboring 
$^{43}$S$_{27}$ nucleus, which will be described in Section~\ref{sectN27}. 

\subsubsection{{\bf $^{42}$Si}} 

\begin{figure}
\centering \epsfig{height=7cm, file=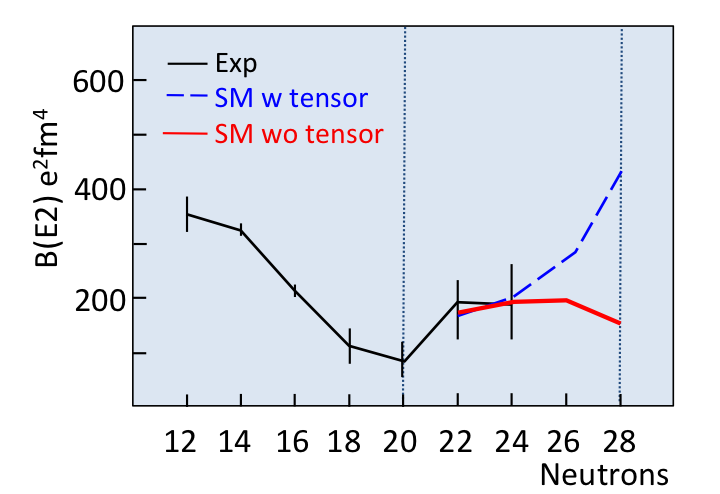} 
\caption{Evolution of the $B(E2)$ values in the Si isotopic chain. 
Experimental values are taken from  Ref.~\cite{Ibbo98} while 
theoretical values are taken from Shell-Model calculations using the sdpf-U interaction~\cite{Nowa09} in which
monopole matrix elements contain tensor effects (dashed blue line)~\cite{Nowacki_priv}. 
The result obtained  when removing this tensor part is shown with a red line. The role of tensor forces is
mainly seen at $N=28$}
\label{BE2inSi}
\end{figure}

As to the $^{42}$Si$_{28}$ nucleus, Bastin \etal~\cite{Bast06} 
have established a $2^+$ state at 770(19)~keV. 
This experiment used nucleon removal reactions from secondary 
beams centered around 
$^{44}$S at intermediate energy to produce the $^{42}$Si 
nucleus and study the $\gamma$-rays from their
de-excitation in flight. The detection of the $\gamma$-rays 
was achieved by arrays of detectors which surrounded the production
target in which the reactions occurred. The dramatic decrease of the 
2$^+_1$ energy in $^{42}$Si is a proof of the disappearance of the 
spherical $N=28$ shell closure at $Z=14$.
This extremely low energy of 770~keV -actually one of the smallest 
among nuclei having a similar atomic mass - cannot be obtained
solely from neutron excitations. Proton-core excitations should 
play an important role, which could in principle be evidenced by
measuring the evolution of the $B(E2)$ values in the Si isotopic chain 
while reaching 
$N=28$. The $B(E2)$ values in the $_{14}$Si isotopic 
chain~\cite{Ibbo98} seem
to rise after $N=20$, but not as much as in the $_{16}$S one (see the
right part of Figure~\ref{E2BE2N2028}). 
Whether the $B(E2)$ values remain small, steadily increase up to
$N=28$, or follow a parabola cannot be judged at the present time 
as the quoted experimental error bars are too large. A reduced 
$Z=14$ shell gap would dramatically 
increase the $B(E2)$ values, as protons are carrying most of the 
effective charge in the nucleus. The sole decrease of
the $N=28$ gap would barely change the 2$^+_1$ and $B(E2)$ values~\cite{Bast06}. 
The effect of the tensor force is to reduce the 2$^+_1$ energy and enhance the 
$B(E2)$ value at  $N=28$ as shown in Figure~\ref{BE2inSi}, leading to 
$B(E2; 0^+_1 \rightarrow 2^+_1)$ $\sim$ 430 e$^2$fm$^4$ in $^{42}$Si. 
Without implementing tensor parts in 
the monopole terms, the $B(E2)$ in $^{42}$Si drops down to 150 e$^2$fm$^4$.
One could deduce that the studies of the 2$^+_1$ state and the $B(E2)$ value 
in $^{42}$Si 
are essential to ascertain the role of tensor forces at $N=28$.
%%%%%%%%%%%%%%%%%%%%%%%%
%%%%%%%%%%%%%%%%%%%%%%%%

To conclude about the role of the tensor term in mean-field models, it requires a word 
of caution. Indeed, the B(E2) value of $^{42}$Si has been calculated in the 
relativistic Hartree-Bogoliubov model of  Li \etal~\cite{Li11} and in the 
Gogny (D1S) model of Rodr\'iguez-Guzm\'an \etal~\cite{Guzm02}. These two 
models do not contain tensor force. However their B(E2) values differ 
significantly: 200 e$^2$fm$^4$ in ~\cite{Li11} and about 470 e$^2$fm$^4$ 
in ~\cite{Guzm02}.  As far as the 2$^+_1$ energy is concerned, the value 
of Ref.~\cite{Li11} is twice as large as the experimental value, while 
the one of Ref.~\cite{Guzm02} is closer to experiment. Therefore the 
$B(E2)$ values should be measured and compared to theory in the \emph{whole} 
Si isotopic chain to see if a significant increase is occurring at $N=28$. 
The evolution of the $sd$ proton orbits should be used as well as they 
influence strongly the 
B(E2) and E($2^+$) values. 

Further discussions on the implementation of 
tensor interaction and its role in the evolution of the gaps could be found 
for instance in Refs.~\cite{Otsuka-present,Otsu05,Otsu10,Smir10}. 

Results of a new experimental study of the excited states of $^{42}$Si 
have just been published~\cite{Take12}. Thanks to the intense radioactive beams
provided at RIKEN RIBF which enable $\gamma-\gamma$ coincidence measurements, the most
probable candidate for the transition from the yrast 4$^+$ state to the 2$^+$ was 
identified, leading to a 4$^+_1$ energy of 2173(14)~keV. Then the energy ratio, 
$R_{4/2} \sim 2.9$, corresponds to a well-deformed rotor. In addition, two other 
$\gamma$ lines were measured at high energy (at 2032(9) and 2357(15)~keV), which 
would deserve to be better characterized in order to assign other excited states
of $^{42}$Si.

\subsubsection{{\bf $^{40}$Mg}} 

Some years ago, the observation of three $^{40}$Mg nuclei in the 
fragmentation of a primary beam of $^{48}$Ca impinging a W target has 
ended the speculation about the location of the neutron drip-line at
$Z=12$~\cite{Baum07}. This isotope is predicted to lie inside the neutron
drip line in many theoretical calculations (see for instance
Ref.~\cite{Erle12} and references therein). The relativistic Hartree-Bogoliubov model for triaxial
nuclei was used to calculate the binding energy map of $^{40}$Mg in the 
$\beta-\gamma$ plane~\cite{Li11}, which predicts that this very neutron-rich
isotope shows a deep prolate minimum at $(\beta,\gamma)=(0.45, 0^{\circ})$. 
Shell model~\cite{Nowa09} as well as Gogny (D1S) ~\cite{Guzm02} calculations 
predict an extremely prolate rotor as well. 
The identification of the first excited state of $^{40}$Mg remains an 
ambitious challenge for the future. Conjectures about shell evolutions below $^{42}$Si
will be provided in Section~\ref{below42Si}.

%%%%%%%%%%%%%%%%%%%%%%%%%%%%%%%%%%%%%%%%%%%%%%%%
\section{Evolution of $sd$ proton orbits as a function of the neutron 
number\label{proton_espe}}

\subsection{The $\pi 2s_{1/2}$ and $\pi 1d_{3/2}$ orbits}\label{protonN28}
\begin{figure}
\centering \epsfig{height=10cm, file=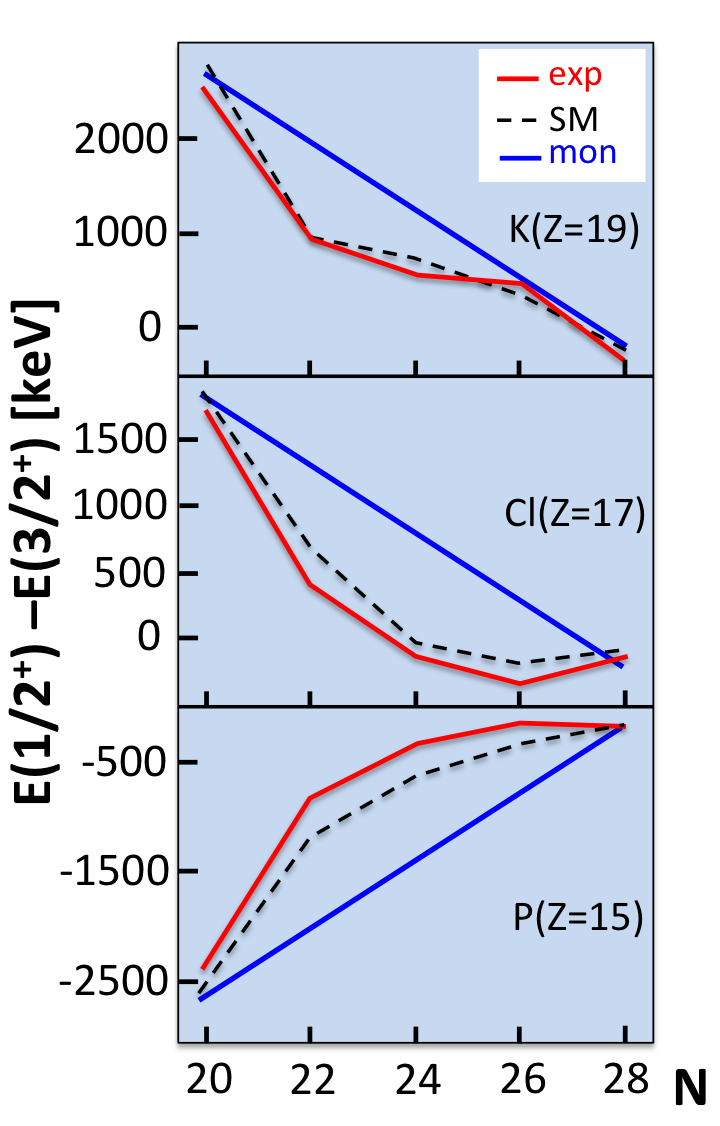} 
\caption{Calculated (black dashed line) and experimental (red line) values of 
$[E(1/2^+_1)-E(3/2^+_1)]$ along the K, Cl and P 
chains (adapted from Ref.~\cite{Gade06}). The fixed monopole-driven 
trend (blue line) given by Equation~\ref{eq:Z16monopole} accounts well for the 
global evolution of $[E(1/2^+_1)-E(3/2^+_1)]$, while correlations 
are required at mid-shell. Note that for the P chain the 
ordering of the $1/2^+$ and $3/2^+$ states has changed as they correspond to particle states 
while they are hole states in the Cl and K chains.}
\label{s1d3inKClP}
\end{figure}

The change of structural behavior between the $N=20$ and 
$N=28$ isotones can be partly ascribed to the evolution 
of the proton
Single-Particle Energies SPE~\cite{Cott98}. Using the pick-up 
reaction ($d$,$^3$He) from stable $_{20}$Ca targets,  
the evolution of the $\pi 2s_{1/2}$ and $\pi 2d_{3/2}$ spacing 
has been revealed in the $_{19}$K isotopic chain.  The filling 
of eight neutrons in the $\nu 2f_{7/2}$ orbital induces an 
enhanced binding energy of the $\pi d_{3/2}$ orbit
as compared to the $\pi 2s_{1/2}$ one. The spacing $\pi 2s_{1/2} -\pi 1d_{3/2}$
derived from SPE values 
drops from 2.52 MeV to -0.29 MeV, i.e. the orbits have crossed 
at $N=28$. This is likely to be due to the fact that the monopole 
interaction $|V{^{pn}_{1d_{3/2} 1f_{7/2}}}|$  is more attractive 
than the $|V{^{pn}_{2s_{1/2} 1f_{7/2}}}|$ one. In Ref.~\cite{nous} 
it was derived that 
%%%%%%%%%%%%%%
\begin{equation}\label{Vsdf}
 V^{pn}_{1d_{3/2} 1f_{7/2}}- V^{pn}_{2s_{1/2} 1f_{7/2}} \simeq -350~keV.
\end{equation}
%%%%%%%%%%%%%% 
The fact that $V{^{pn}_{1d_{3/2} 1f_{7/2}}}$ is significantly 
more attractive than $V{^{pn}_{2s_{1/2} 1f_{7/2}}}$ could 
qualitatively be ascribed to the fact that the proton 
1$d_{3/2}$ ($\ell$=2, $\ell \downarrow$) and neutron 
1$f_{7/2}$ ($\ell$=3, $\ell \uparrow$) wave functions 
have the same number of nodes and have attractive tensor 
terms as the proton  and neutron spins are anti-aligned. On the other hand 
for $V{^{pn}_{2s_{1/2} 1f_{7/2}}}$, the numbers of nodes 
differ and a large difference in the orbital momentum value is present between the two wave functions , 
making this monopole term weaker. 
Taking into account the monopole matrix elements solely, 
the evolution of [$E(1/2^+)-E(3/2^+)$] between $N=20$ 
and $N=28$ would be linear as a function of the number 
of neutrons $x$ leading to:
%%%%%%%%%%
\begin{equation}\label{eq:Z16monopole}
[E(1/2^+)-E(3/2^+)]_{20+x}=[E(1/2^+)-E(3/2^+)]_{20}
+x(V{^{pn}_{d_{3/2}f_{7/2}}}-V{^{pn}_{s_{1/2}f_{7/2}}})
\end{equation}
%%%%%%%%%%
The experimental evolution of [$E(1/2^+_1)-E(3/2^+_1)$] 
deviates at mid shell from the linear monopole trend
shown in Figure~\ref{s1d3inKClP}. This deviation is due to 
pairing and quadrupole correlations which already 
engage at $N=22$, as
soon as the single-particle states $\pi s_{1/2}$ and 
$\pi d_{3/2}$ come close enough to each other. 
Shell model calculations using the $sdpf$ interaction 
\cite{Numm01a} well reproduce these correlations~\cite{Sorl04,Gade06}, 
as shown by the black dashed line in Figure~\ref{s1d3inKClP}. 
%%%%%%%%%%%%%%%%%%%%
%%%%%%%%%%%%%%%%%%%%

The strong reduction between the 
proton $3/2^+$ and $1/2^+$ states is also found around $N=28$ 
in the $_{17}$Cl and $_{15}$P isotopic chains.
This demonstrates that the change in monopole interaction 
plays a decisive role in bringing the $\pi s_{1/2}$ and 
$\pi d_{3/2}$ states degenerate at $N=28$.
This has a profound consequence on the evolution of collectivity 
between $N=20$, where a sub-shell gap exists between 
$\pi s_{1/2}$ and $\pi d_{3/2}$, and $N=28$, 
where this sub-shell has vanished completely.  As discussed above
for the K isotopes, the evolution of the experimental $3/2^+_1$ and 
$1/2^+_1$ states for the Cl and P isotopes is
distorted by pairing and quadrupole correlations 
(see Figure~\ref{s1d3inKClP}), which are also 
well accounted for by shell-model calculations ~\cite{Gade06}.

\subsection{The $\pi d_{5/2}$ orbit}\label{protond5N28}

As shown before, the evolution of the $Z=14$ shell gap is crucial for providing enhanced correlations
in the $N=28$ nuclei far from stability, as well as to probe the effect of tensor forces. However, as the single-particle 
strength is significantly spread in the $^{43}$P nucleus,  the size of the $Z=14$ gap can hardly be extracted there from experimental data.
In order to extract more accurately the change in the size 
of the $Z=14$ gap when filling the neutron $f_{7/2}$ orbit 
from $N=20$ to $N=28$ we could in principle look at the evolution of 
the $s_{1/2}$ and $d_{5/2}$ single-particle energies in the 
K isotopic chain as deformation is not developing there. 

\begin{figure}
\centering \epsfig{height=8cm, file=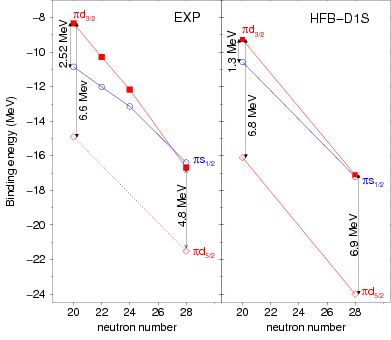} 
\caption{Left: Evolution of the proton $d_{5/2}$, $s_{1/2}$ and 
$d_{3/2}$ orbits in the K isotopic chain derived from experimental 
data. Right: Results of HFB calculations using the 
Gogny D1S interaction (see text).}
\label{espeK}
\end{figure}

However, one additional problem arises from the fact that in the K chain 
the $d_{5/2}$ orbit appears to be more bound by about 2.5~MeV 
than in the P isotopic chain. It follows that it is very hard to study 
the evolution of the $d_{5/2}$ single particle energy in 
the K chain as the  $d_{5/2}$ strength is spread into many 
states which carry a small fraction of it 
(see the discussion in Section~4.2.3 of Ref.~\cite{nous} 
and its Tables 2 and 3). 
With this important word 
of caution in mind, we can nevertheless discuss the evolution of the binding energies of the $\pi d_{5/2}$, 
$\pi s_{1/2}$ and $\pi d_{3/2}$ orbits between $^{39}$K$_{20}$ and 
$^{47}$K$_{28}$ in a semi quantitative way. The experimental part of Figure~\ref{espeK} 
displays a reduction of the proton $d_{5/2} - d_{3/2}$ 
splitting by about 1.7 MeV between $N=20$ and $N=28$.
%%%%%%%%%%%%%%%%%%% 
%%%%%%%%%%%%%%%%%%%
When adding neutrons in the $f_{7/2}$ shell, the two $d$ orbits 
become more bound. This is mainly due to the attractive proton-neutron 
interactions. The fact that the gain in 
binding energy is larger for the $1d_{3/2}$ orbit than 
for the $1d_{5/2}$ one comes from the fact that the 
$V{^{pn}_{1d_{5/2} 1f_{7/2}}}$ monopole is weaker than $V{^{pn}_{1d_{3/2} 1f_{7/2}}}$.  
It likely comes (at least partly) from the fact that 
$V{^{pn}_{1d_{3/2} 1f_{7/2}}}$ ($V{^{pn}_{1d_{5/2} 1f_{7/2}}}$) 
monopole contains an attractive (repulsive) tensor part 
as the proton and the neutron have anti-aligned (aligned) 
spin orientations. 

Note that the size of the $Z=14$ gap, which is almost similar for the two isotones 
$^{39}$K and $^{35}$P (the properties of $^{35}$P are discussed in Section~\ref{phosphore}), 
is an essential ingredient for 
providing a similar structural behavior at $N=20$ from $^{40}$Ca to $^{34}$Si. 

\subsection{Theoretical predictions of the evolution of the 
three orbits in the K isotopic chain}\label{protons_th_N28}

Interesting is to compare this variation of proton orbits to calculations 
obtained with Hartree-Fock (HF) self-consistent 
calculations using the D1S Gogny force~\cite{D1S,D1Sbis}.  
The binding energies of the single-particle levels of 
$^{40}$Ca and $^{48}$Ca have been obtained
from HF calculations. As usual, the HF equations are firstly solved by
iterative diagonalization of the HF Hamiltonian. Then the single-particle
energies are defined as the eigenvalues of the self-consistent one-body
Hamiltonian, obtained after convergence.
In this approach, a change of the $s_{1/2} - d_{3/2}$ 
splitting is found from $N=20$ to $N=28$, but twice as small as experimentally. No change of the $d_{5/2} - d_{3/2}$ 
splitting is obtained (see the right part of Figure~\ref{espeK}) as tensor interaction is not included. Moreover, as these orbits 
are well bound no reduction of the spin-orbit 
interaction associated with an extended surface density profile is foreseen. We conclude that a further reduction of the $d_{5/2} - d_{3/2}$ splitting is required to match the experimental results.

\subsection{Comparison of $^{35}$P$_{20}$ and $^{43}$P$_{28}$}
\label{phosphore}
The development of collectivity in the $_{16}$S and $_{14}$Si 
isotopic 
chains depends on the amount of proton excitations across 
the $Z=14$ gap,
from the deeply bound proton $d_{5/2}$ to the degenerate 
$d_{3/2}$ and $s_{1/2}$ orbits (see the bottom part of Figure~\ref{Fig1}). 
When possible, the  $s_{1/2}$-$d_{5/2}$ cross-shell excitations 
naturally bring quadrupole excitations. The comparative studies of 
the $^{35}$P$_{20}$ and $^{43}$P$_{28}$ nuclei gives valuable information
on the reduction of the $Z=14$ gap and the increasing role of 
correlations across it. 

\begin{figure}
\centering \epsfig{height=10cm, file=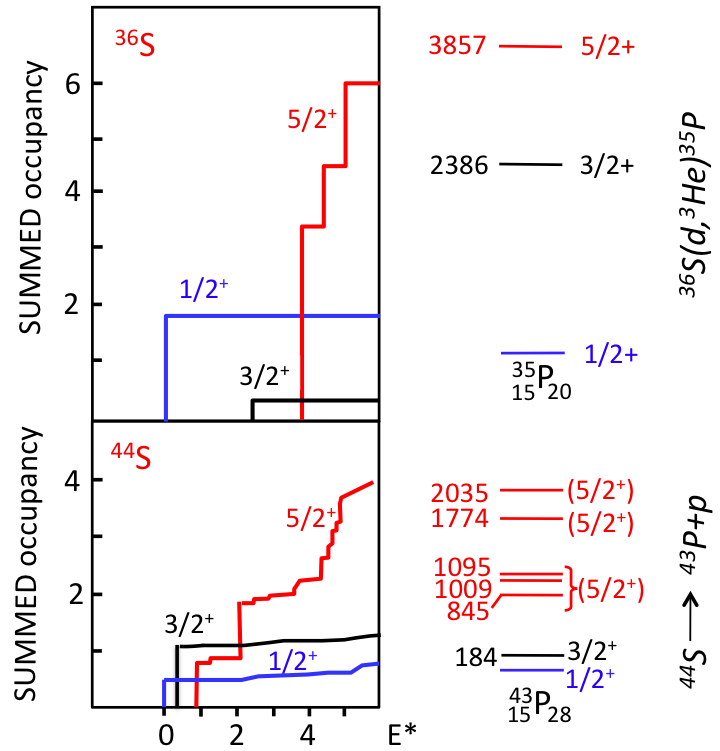} 
\caption{Top: Occupancy values of the proton $d_{5/2}$, 
$s_{1/2}$ and $d_{3/2}$ orbits are given for $^{35}$P$_{20}$. 
They were derived  from spectroscopic factors obtained in 
the $^{36}$S$_{20}$(d,$^3$He)$^{35}$P$_{20}$ reaction~\cite{Khan85}. 
The first excited states of $^{35}$P$_{20}$ are shown in 
the right part of the figure. Bottom (adapted from Ref.~\cite{Riley08}): calculated occupancy values
for $^{43}$P$_{28}$  derived from the $^{44}$S$_{28}$ (-1p) $^{43}$P$_{28}$ reaction. 
The level scheme of $^{43}$P$_{28}$ shown in 
the right part of the figure contains many $5/2^+$ states, 
witnessing a large spreading of the proton $d_{5/2}$ strength 
in this nucleus.}
\label{d5inP}
\end{figure}

The energy of the three orbits, $\pi d_{5/2}$, $\pi s_{1/2}$ and $\pi d_{3/2}$,  
and their occupancy values, $(2J+1) \times SF$ (where $SF$ is the spectroscopic factor), 
have been obtained in $^{35}$P$_{20}$ by Khan \etal~\cite{Khan85} 
by means of  the $^{36}$S$_{20}$($d$,$^3$He)$^{35}$P$_{20}$
reaction. The major results are shown in the top part of 
Figure~\ref{d5inP}. 
%%%%%%%%%%%%%%%%%%%%%
%%%%%%%%%%%%%%%%%%%%%
The energy spectrum of $^{35}$P$_{20}$ 
exhibits few levels, with rather large spacing.
The sum of the 
occupancies of the two first states $s_{1/2}$ 
and $d_{3/2}$  is about 2, while most of the occupancy is 
in the $s_{1/2}$ orbit. As the $s_{1/2}- d_{3/2}$ spacing is not 
large enough (about 2.4 MeV) to  smother pair scattering,
some leak of occupancy from the $s_{1/2}$ to the upper $d_{3/2}$ 
orbit is present.  The full $d_{5/2}$ 
strength is shared in three major fragments between 3.8 
and 5.2 MeV excitation energy, leading to a mean value 
of the $Z=14$ gap of 4.6 MeV. The picture is there rather 
simple, with a  sequence of $d_{5/2}$,  $s_{1/2}$ and $d_{3/2}$ 
orbits bound by about 16.8 MeV, 12.2 and 9.8 MeV, respectively.

The situation gets more complex in the $^{43}$P$_{28}$ nucleus, 
which was studied by means of one proton knock-out reaction 
$^{44}$S$_{28}$ (-1p) $^{43}$P$_{28}$ by Riley \etal~\cite{Riley08}. 
At first glance it is clear in Figure~\ref{d5inP} that the energy 
spectrum of $^{43}$P$_{28}$ is more compressed and contains 
many more levels than $^{35}$P$_{20}$ does. This compression of levels is 
relatively well reproduced by shell-model calculations using the interaction of 
Utsuno \etal~\cite{Utsu07}. If the sum of 
the $s_{1/2}$ and $d_{3/2}$ occupancies still leads to about 
2, the sharing of occupancy reflects the quasi degeneracy 
of these two orbits: since the $d_{3/2}$ orbits contains twice 
as many sub-states as the $s_{1/2}$ does its occupancy is 
implicitly twice as large. As to the $d_{5/2}$ strength, 
it is spread over many states among which only the ones below
2 MeV have been observed experimentally. Noteworthy is the 
fact that the $d_{5/2}$ strength already starts at low energy 
and a significant fraction is already calculated below 3 MeV. 
This high density of states at low energy as well as the 
spreading of the occupancy value are at variance with the 
picture observed in $^{35}$P. In order to obtain a significant 
$d_{5/2}$ strength below 3.5 MeV in $^{43}$P, the spherical $Z=14$ 
gap has to be slightly reduced between $N=20$ and $N=28$~\cite{Utsu07,Bast06}. 

\subsection{Summary}
The development of collectivity in the 
neutron-rich $N=28$ isotones is partly due to the reduction of the
spacings between the protons $d_{5/2}$, s$_{1/2}$ and $d_{3/2}$ orbits
as soon as the neutron $f_{7/2}$ orbit is completely filled. As 
these orbits are
separated by 2 units of angular momentum, quadrupole (E2) 
collectivity is naturally favored. Added to this, a reduction
of the $N=28$ shell gap would reinforce this tendency to deform. Mean-field
and shell-model theories agree on this description. We note that from the evolution of the proton $sd$ 
orbits, mean-field theories need an additional spin-dependent term to 
reduce the $d_{5/2}- d_{3/2}$ splitting in order to match the experimental value.

It is important to note that a systematic study~\cite{Gaud08,Gaud10} of the 
low-lying structures of many isotopic and isotonic chains was performed 
within the shell model approach, using the interaction of Nowacki and Poves~\cite{Nowa09}. In these works, a particular focus is also made on the
cases with unpaired nucleons. Indeed
the properties of the $N=27$ and $N=29$ isotones give a deeper understanding
of the evolution of the $N=28$ shell gap, as detailed in the two next
Sections.

\section{Structural evolution viewed from the $N=27$ isotones.}
\label{sectN27}

\begin{figure}
\centering \epsfig{height=11cm, file=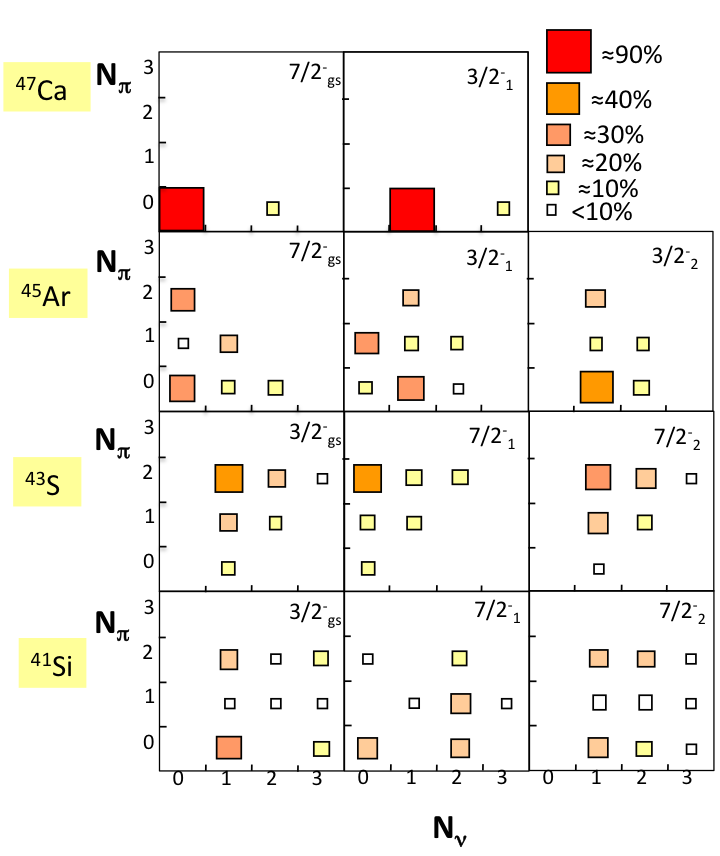} 
\caption{Squared wave functions of the first $7/2^-$ and $3/2^-$ states in the
$N=27$ isotones
represented in the  proton ($N_\pi$) versus neutron ($N_\nu$) particle-hole
configuration plane (see text for details), adapted from Ref.~\cite{Gaud08}.
Some of the excitation energies are given in the text.}
\label{N27}
\end{figure}

The possible development of collectivity can be looked at in 
the $N=27$ isotones, by comparing the characteristics of their first 
states. Shell-model calculations~\cite{Gaud08} have been performed 
by Gaudefroy~\etal using the interaction of Nowacki and Poves~\cite{Nowa09}. The results are drawn in 
Figure~\ref{N27}. In this plot, the configurations 
of the ground and the first excited states are displayed using
a bi-dimentional representation, where the proton (neutron) 
configuration is shown on the y (x) axis. 
$N_\pi$ ($N_\nu$) gives the number of proton (neutron) \emph{excitations} 
above the $0p0h$ ($0p1h$) core configuration and the size of the squares 
in the $N_\pi- N_\nu$ representation gives the intensity of 
each component. 
%%%%%%%%%%%%%%%%%%%%%%%%%%%%%%%%%%%
%%%%%%%%%%%%%%%%%%%%%%%%%%%%%%%%%%%

About 90\% of the g.s. configuration of the spherical $^{47}$Ca nucleus
corresponds to a neutron hole inside the $f_{7/2}$ shell ($0p1h$, $N_\nu$=0) 
and a closed proton core ($0p0h$, $N_\pi$=0).
The first-excited state with $I^\pi=3/2^-$ at 2.02~MeV is expected to 
involve the promotion of 
one neutron in the upper $p_{3/2}$ shell with two neutron holes coupled 
in the $f_{7/2}$ shell ($1p2h$, $N_\nu$=1). This is in agreement with 
the theoretical results showing that the first-excited state at 2.07~MeV 
corresponds mainly to the pure excitation 
of one neutron ($N_\nu$=1) and to a minor extent, of three neutrons ($N_\nu$=3).
This state, with a closed proton configuration, would be 
found well above the ground state of all the $N=27$ isotones, 
except if the $N=28$ shell gap is reduced and correlations dominate.

The configurations of the g.s. and excited states in $^{45}$Ar  
have been studied using $^{44}$Ar(d,p)$^{45}$Ar~\cite{Gaud08} 
and $^{46}$Ar(-1n)$^{45}$Ar~\cite{Gade05} transfer and knock out 
reactions, respectively. Very good agreement is found between
the experimental results, excitation energies and spectroscopic factors, 
and the theoretical predictions~\cite{Gaud08}. The drawings of 
Figure~\ref{N27} show that as compared 
to  $^{47}$Ca into which proton excitations are hampered by 
the $Z=20$ gap, proton excitations are naturally present in $^{45}$Ar inside 
the $s_{1/2}d_{3/2}$ proton states which are degenerate in 
energy at $N \sim 28$, as discussed in Section~\ref{protonN28}. These 
proton excitations occur as $1p1h$ ($N_\pi=1$) or $2p2h$ ($N_\pi=2$). They 
both contribute to the largest fraction of the $B(E2)$ values in the 
$Z=16,18$ nuclei. In line with these features the $^{45}$Ar ground state is still dominated by a 
configuration 
similar to the one of $^{47}$Ca, but correlations leads to a more 
mixed wave function. The state corresponding to the first-excited 
state in $^{47}$Ca is more likely the second-excited state 
in $^{45}$Ar ($3/2^-_2$) observed at 1.42~MeV and predicted at 1.22~MeV, 
for which the $N_\nu$=0 component 
does not exist either. The configuration of the ($3/2^-_1$) 
state at 550 keV is in between the $7/2^-_{gs}$ and the  $3/2^-_2$ configuration, 
which already witnesses the sharing of the $p$ strength 
among several states.  

Below $^{45}$Ar, an inversion between the $7/2^-$ and $3/2^-$  states
is predicted (see Figure~\ref{N27}).  An isomer has been discovered 
in $^{43}$S by Sarazin \etal at 320 keV~\cite{Sara00}. It was 
interpreted with the help of SM calculation to have a  
$7/2^-$ spin value and to decay to the $3/2^-$ ground state by a 
delayed transition. More recently, the $g$ factor
of this isomer was measured~\cite{Gaud09} establishing its spin-parity
values, $I^\pi=7/2^-$, and implying a rather spherical shape for this state. 
Nevertheless, the value of its quadrupole moment is larger than expected for a
single-particle state~\cite{Chev12}, implying that it contains non-negligible 
correlations that drive the state away from a purely spherical shape, 
in agreement with SM calculations (see Figure~\ref{N27}). 
In addition, an intermediate-energy single-neutron knockout
reaction was used to characterize other excited states of $^{43}$S
~\cite{Rile09b}. Two of them are proposed to be members of the rotational 
band built on the deformed $3/2^-$ ground state, strengthening the case 
for shape coexistence in $^{43}$S. 

Note that this inversion of the $7/2^-$ and $3/2^-$ states, and the spreading of the wave 
function among neutron and proton excitations is, according 
to SM calculations, persisting in the $^{41}$Si$_{27}$ isotope.
The spectroscopy of $^{39,41}$Si has been carried out by 
Sohler \etal~\cite{Sohl11}  using the in-beam $\gamma$-ray
spectroscopy method from few nucleon knockout reactions.  The 
observation of low lying states in
$^{39}$Si call for a reduction of the $N=28$ and $Z=14$ 
shell gaps
which induce the lowering of the intruder neutron 3/2$^-$ state 
while going from $Z=20$ to $Z=14$. The energy of the only
$\gamma$ line, 672(14)~keV, observed in $^{41}$Si is significantly 
lower than the one of the first excited
state in $^{47}$Ca (2014~keV) suggesting a deformed ground 
state for $^{41}$Si. However, as compared to shell-model predictions, some 
states expected at low energy have not been observed in $^{41}$Si. It 
was suggested in Ref.~\cite{Sohl11} 
that a low-energy isomer would be present there. As isomers 
could not be evidenced with in-beam spectroscopy, another technique 
has to be used to reveal its existence. The search for this isomer and/or low-energy states 
would bring important pieces of information to confront to SM description. In Figure~\ref{N27} one
can observe that for $^{41}$Si many squares have equal size, pointing to large mixing of proton and neutron configurations.

As mentioned above, the inversion between the natural 
$(\nu f_{7/2})^{-1}$ and the intruder $(\nu p_{3/2})^{+1}$ configurations
of the ground states of the $N=27$ isotones
occurs between $^{45}$Ar and $^{43}$S. The properties of the ground state 
of $^{44}$Cl therefore allow us to delineate the exact
location of the inversion. This odd-odd nucleus has been populated by using the single
neutron-knockout reaction from $^{45}$Cl$_{28}$ at
intermediate beam energies at MSU~\cite{Rile09a}. The momentum distribution of the
recoiling $^{44}$Cl nuclei was analyzed for the direct population of its
ground state, the best fit being obtained for the removal of an $\ell=1$
neutron. This means that (i) there is a significant $(\nu
p_{3/2})^{+2}$ component in the ground state of $^{45}$Cl and (ii) 
this neutron orbit is involved in the ground state of $^{44}$Cl.
Therefore the intrusion of the $\nu p_{3/2}$ orbit from above the $N=28$
shell closure occurs already for $Z=17$. This is confirmed by the value of
the $g$ factor of the ground state of $^{44}$Cl. This measurement, done
at the LISE fragment separator at Ganil using the $\beta$ nuclear magnetic
resonance technique~\cite{Rydt10}, 
indicates that $g$($^{44}$Cl) is significantly lower than $g$($^{46}$K). 
The two $g$ values are well reproduced by SM calculations. 
While the main configuration of the ground state of $^{46}$K is
dominated by the spherical $(\pi d_{3/2})^3(\nu f_{7/2})^7$ configuration,
the wave function of the ground state of $^{44}$Cl is very fragmented,
its second most-intense component, $(\pi s_{1/2})^1(\nu p_{3/2})^1$, 
explaining why the $g$ factor of $^{44}$Cl is reduced~\cite{Rydt10}.

%%%%%%%%%%%%%%%%%%%%%%%%%%%%%%%%%%%
\section{Structural evolution viewed from the $N=29$ isotones.}
\label{neutron_espe}
\begin{figure}[t]
\begin{center}
\includegraphics[width=12cm]{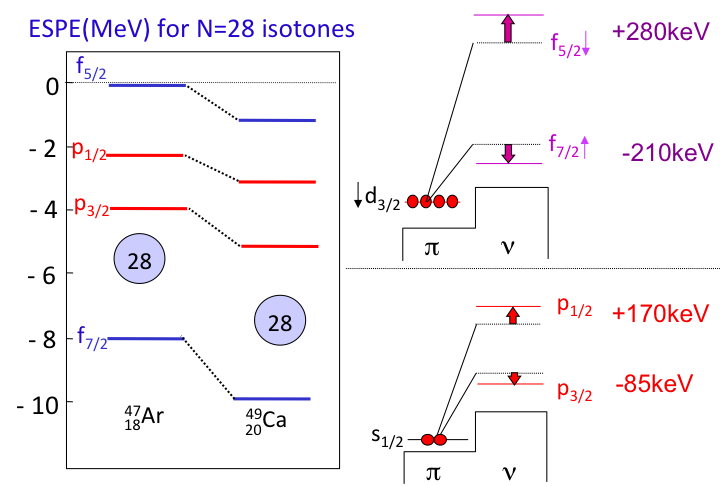}
\end{center}
\begin{center}
%\begin{minipage}{16.5cm}
\caption{Left: Neutron single-particle energies (SPE) of the $fp$ orbitals for the
$^{47}_{18}$Ar$_{29}$ and $^{49}_{20}$Ca$_{29}$ nuclei.
Right: Schematic view of the proton-neutron spin-dependent  interactions
involved to change the $f$ (top) and $p$ (bottom) SO splittings
derived from the experimental data on $^{47}$Ar in Refs.~\cite{Gaud06,Gaud07}.} 
\label{spe_N28}
%\end{minipage}
\end{center}
\end{figure}

%%%%%%%%%%%%%%%%%%%%%%%%%%%%%%%%%%%
In the previous sections, the erosion of the $N=28$ shell gap below the doubly magic $^{48}_{20}$Ca
has been probed through properties of atomic masses, energies 
of excited states and reduced transition probabilities, as well as through the spectroscopy of 
$N=27$ isotones.  The most direct way to evidence this erosion is to determine the 
evolution of the single particle energies between the 
$^{49}_{20}$Ca and $^{47}_{18}$Ar isotones. The study of  $^{47}$Ar$_{29}$ has been carried out by the
$^{46}_{18}$Ar($d,p$) transfer reaction in inverse 
kinematics at the GANIL/SPIRAL1 facility~\cite{Gaud06}. The obtained results were used to
investigate the change of the $N=28$ gap from $_{20}$Ca to 
$_{18}$Ar. From the $Q$ value of the transfer
reaction, the $N=28$ gap was found to be of 4.47(9)~MeV in 
$^{46}$Ar, which is 330(90)~keV smaller than in $^{48}$Ca.
Transfer to excited states were used to determine the energies 
and spectroscopic factors of the neutron $p_{3/2}$, $p_{1/2}$ 
and $f_{5/2}$ states in $^{47}_{18}$Ar.
The fact that only part (about 3/4) of the strength for these 
states has been observed indicates
that some correlations takes place already in $^{46}$Ar, after the removal 
of only 2 protons from the doubly magic $^{48}$Ca nucleus. In
particular it was found in Ref.~\cite{Gaud06} that the $p_{3/2}$ orbit
is already partially occupied 
through $1p-1h$ excitation across the $N=28$ gap, from the 
$f_{7/2}$ orbit. The agreement between experiment results and SM calculations was rather good using 
the monopole terms of the $sdpf$~\cite{Numm01a}
interaction. They were however slightly  adjusted  to match existing 
experimental data in this mass region~\cite{Gaud07}. By applying
this technique, the full particle strength of the neutron $f_{7/2}$, 
$p_{3/2}$ and $p_{1/2}$ and $f_{5/2}$ orbits has been
determined in $^{47}$Ar. The resulting SPE
 are compared to those of the $^{49}_{20}$Ca
isotone~\cite{Abeg78,Uozu94a} in Figure~\ref{spe_N28}. It is found 
that from $^{47}$Ar to $^{49}$Ca the orbits in which the angular momentum is aligned 
with the intrinsic spin ($\ell_\uparrow$), such as $f_{7/2}$ 
and $p_{3/2}$, become relatively more bound than the $f_{5/2}$
and $p_{1/2}$ orbits where the angular momentum and intrinsic 
spin are anti-aligned ($\ell_\downarrow$). In particular, one can see in Figure~\ref{spe_N28} the asymmetry
in gain of binding energy between the $f_{7/2}$ and $f_{5/2}$ orbits.

Bearing in mind that 
the $d_{3/2}$ and $s_{1/2}$ orbitals are quasi-degenerate at $N=28$, 
(see Section~\ref{protonN28}), the addition of  2 protons between 
$^{47}$Ar and $^{49}$Ca occurs in an equiprobable manner\footnote{When
taking into account the correlations expected in non doubly-magic nuclei, the
occupation rates of the orbits are slightly modified, see for instance the 
numbers given in Ref.~\cite{Gaud07}} in these orbits 
(i.e 1.33 in $1d_{3/2}$ and 0.66 in $2s_{1/2}$). 
Therefore modifications of neutron SPE arise from proton-neutron
interactions involving these two orbits~\cite{Gaud06} and
the change of the $N=28$ shell  gap ($\delta G$) can be approximated to :
%%%%%%%%%%%%%%
\begin{equation}\label{GAP28}
\noindent \delta G^{pn}(28) \simeq 1.33(V_{1d_{3/2} 2p_{3/2}}^{pn} - V_{1d_{3/2}
1f_{7/2}}^{pn}) + 0.66(V_{2s_{1/2} 2p_{3/2}}^{pn} - V_{2s_{1/2}
1f_{7/2}}^{pn})
\end{equation}
%%%%%%%%%%%%%%
Similarly,  changes in the $p$ and $f$ SO splitting express as:
%%%%%%%%%%%%%%
\begin{equation} \label{SO}
\delta SO= 1.33(V_{1d_{3/2} \ell_{\downarrow}}^{pn} - V_{1d_{3/2}
\ell_{\uparrow}}^{pn}) + 0.66(V_{2s_{1/2} \ell_{\downarrow}}^{pn} -
V_{2s_{1/2} \ell_{\uparrow}}^{pn})
\end{equation}
%%%%%%%%%%%%%%
As regards the $f$ states, the experimental change in SO splitting 
$\delta SO(f)$ was ascribed in \cite{Gaud06} to the fact that the monopoles 
contain attractive and repulsive spin-dependent terms (which could be tensor 
terms) :
%%%%%%%%%%%%%%
\begin{equation} \label{SOf}
\delta SO(f) \simeq 1.33(V_{1d_{3/2} 1f_{5/2}}^{pn} - V_{1d_{3/2} 1f_{7/2}}^{pn}) 
\end{equation}
%%%%%%%%%%%%%%
As shown in Figure~\ref{spe_N28} the spin-dependent  parts of the monopoles $\tilde{V}$ amount to 
$\tilde{V}_{1d_{3/2}~ 1f_{7/2}}^{pn}$ =-210~keV and
$\tilde{V}_{1d_{3/2} ~1f_{5/2}}^{pn}$=+280~keV, respectively. They  amount to 
about 20\% of the total monopole term $V_{1d ~1f} ^{pn}$.  The change of the $p$ SO splitting 
$\delta SO(p)$ was principally assigned in \cite{Gaud07} to the
removal of a certain fraction of $2s_{1/2}$ protons which depletes the central 
density\footnote{Having an angular momentum $\ell$=0, the $s_{1/2}$ orbit 
is peaked in the center of the nucleus. When no proton occupies the 
$s_{1/2}$ orbit, the central density is therefore depleted as compared 
to nuclei in which this orbit is filled. This property will be used 
in Section~\ref{SOtensor} to study the density dependence of the SO 
interaction.} of the nucleus:
%%%%%%%%%%%%%%
\begin{equation} \label{SOp}
\delta SO(p) \simeq  0.66(V_{2s_{1/2} 2p_{1/2}}^{pn} -
V_{2s_{1/2} 2p_{3/2}}^{pn})
\end{equation}
%%%%%%%%%%%%%%
As shown in Figure~\ref{spe_N28}, the corresponding
spin-dependent part of the monopole terms was extracted to be 
of +170~keV and -85~keV for $\tilde{V}_{s_{1/2}~p_{1/2}}^{pn}$ and
$\tilde{V}_{2s_{1/2}~2p_{3/2}}^{pn}$, respectively. 

%%%%%%%%%%%%%%%%%%%%%%%%%%%%%%%%%%%%%%%%%%%%%%%%%%%%%%%%%%
%%%%%%%%%%%%%%%%%%%%%%%%%%%%%%%%%%%%%%%%%%%%%%%%%%%%%%

The variation of the single-particle energies in terms of monopole
interactions could be pursued towards the $^{43}_{14}$Si nucleus in
which, as compared to $^{49}_{20}$Ca, \emph{about} 4 protons have been removed
in the $d_{3/2}$ and 2 in the $s_{1/2}$ orbits. We note here that protons are partly removed in
the $d_{5/2}$ orbit as well. Using the
monopole matrix elements derived in the interaction, 
changes in the spacing of the $fp$ orbits is foreseen when the 
$d_{3/2}$ and $s_{1/2}$ orbits are empty. This leads to  reductions of
(i) the $N=28$ gap by $0.33 \times 3 \simeq 1$~MeV, (ii) the $f$ SO
splitting by $ \simeq4(0.28+0.21) \simeq 2$~MeV and (iii) the $p$ SO
splitting by $ \simeq 2(0.17 + 0.085) \simeq 0.5$~MeV. These global
reductions are expected to reinforce particle-hole excitations
across $N=28$, which are of quadrupole nature. 

It is important to note that, in the shell-model approach, the monopole 
terms do not vary when approaching the drip-line. Therefore when 
reaching the continuum the present approach  and the concepts that 
have been used will be inappropriate. The proximity of 
continuum could already modify the binding energy of the $f_{5/2}$ 
orbit in  $^{47}$Ar, hereby possibly altering the role of tensor 
interaction derived from the present experimental data. The role of continuum should be quantified
in this region.

\section{Questions and perspectives}
%%%%%%%%%%%%%%%%%%%%%%%%%%%%%%%%%%%
\subsection{Introduction}

As described in the previous sections, a wealth of spectroscopic 
information has been obtained from the study of the $N \simeq 28$ 
nuclei from $Z=20$ down to $Z=12$. Several important questions related 
to nuclear forces arose from these studies. In the following we propose 
to address some of them:
\begin{itemize}
\item What is the role of the three-body forces in providing the SO shell 
gaps ? Is there a general picture that is present throughout the chart of 
nuclides ?
\item Can we disentangle the respective roles of the tensor and the spin-orbit  
components to account for the disappearance of $N=28$ far from stability? 
Can we better understand the spin-orbit force ?
\item Would other SO magic numbers (14, 50, 82 and 126) be disappearing 
far from stability as $N=28$ does and if not why ?
\item Which forces come into play in nuclei located below $^{42}$Si, 
$^{78}$Ni and $^{132}$Sn?  
\item Are proton-neutron forces significantly modified when approaching 
the drip-line ?  
\end{itemize}
We propose to discuss these questions using rather qualitative and 
phenomenological arguments based on experiments results.

\subsection{What is the role of the three-body forces in providing the SO shell 
gaps ?}

\subsubsection{{\bf Introduction}}

In this Section we investigate the role of neutron-neutron interactions to create the neutron SO shell gaps. 
The three-body forces are needed to create these gaps, which would not be obtained if realistic two-body forces would be used exclusively.
We start with the study of the $N=28$ shell gap and generalize
this study to other SO shell gaps as $N=14$, $N=50$ and $N=90$. We conclude that 
the three-body forces seem to play a similar and essential role to create the major shell gaps that leads to SO magic nuclei.
We propose an empirical rule to predict the evolution of high-$j$ orbits in superheavy nuclei as well as to create 
a sub-shell gap at $N=90$, which could be of importance for the r-process nucleosynthesis.

\subsubsection{{\bf Neutron-neutron forces to create the $N=28$ shell gap in Ca}}

\begin{figure}[t]
\begin{center}
\includegraphics[width=16cm]{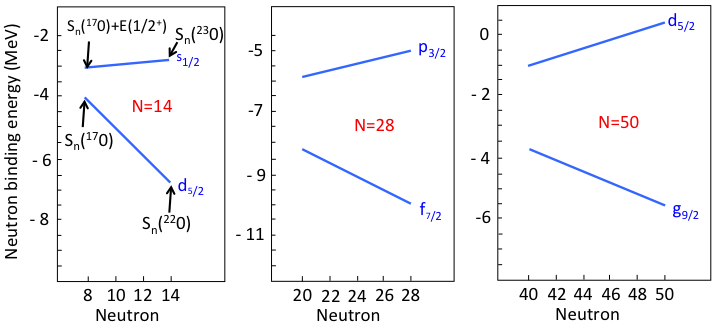}
\end{center}
\begin{center}
%\begin{minipage}{16.5cm}
\caption{Evolution of neutron binding energies in the O (left), Ca (middle) 
and Ni (right) isotopic series. In each chain, the value of the neutron shell 
gap grows when adding neutrons in the high-$j$ orbit ($d_{5/2}, f_{7/2}, 
g_{9/2}$ respectively).  In the cases of  the O and Ca chains, experimental 
values are taken at the two extremes of the binding energy values and a 
straight line is drawn in between, according to the monopole-driven trend.  As for the Ni chain, details are found 
in the text.} 
\label{3body}
%\end{minipage}
\end{center}
\end{figure}

Search for  $fp$ single-particle have been carried out using transfer reactions~\cite{Uozu94b,Uozu94a} at 
$N=20$ and $N=28$ by studying the particle and hole strengths around the 
$^{40}$Ca and $^{48}$Ca nuclei, respectively. 
These experiments show that the $N=28$ shell gap strongly increases when 
\emph{adding neutrons} into the $f_{7/2}$ orbit, as the neutron $f_{7/2}$ ($p_{3/2}$) 
orbit gains (looses) binding energy. The gain in energy of 
the $N=28$ gap from $N=20$ to $N=28$ is about $\delta G^{nn}(28)=3$~MeV (see the middle of Figure~\ref{3body}). 
%%%%%%%%%%%%%%%%%%%%%%%%%%%%%%%%%%%%%%%%%%%%%%%%%%%%%%%%%%
%%%%%%%%%%%%%%%%%%%%%%%%%%%%%%%%%%%%%%%%%%%%%%%%%%%%%%
The mechanism 
which creates the large $N=28$ shell gap, between the  
$f_{7/2}$ and $p_{3/2}$ 
orbits, is likely due to a strongly attractive $V_{f_{7/2}~f_{7/2}}^{nn}$ 
and a repulsive $V_{f_{7/2}~p_{3/2}}^{nn}$ monopole term\footnote{We note here
that the $V^{nn}$ are \emph{effective} monopole terms. Even though they are written 
in a two-body form, they may contain three-body components.}. The increase 
of the $N=28$ shell gap, $\delta G^{nn}(28)$, due to the neutron-neutron 
interactions can be written as a function of x (the number of 
neutrons added in the $f_{7/2}$ orbit) as :
%%%%%%%%%%%%%%%%%
\begin{equation}\label{GAP3body}
\delta G^{nn}(28)=  x [V^{nn}_{f_{7/2}p_{3/2}}] - (x-1) [V^{nn}_{f_{7/2}f_{7/2}}] 
\end{equation}
%%%%%%%%%%%%%%%%%
We note that the $(x-1)$ term applies when neutrons occupy the same shell.

Effective two-body interactions derived from \emph{realistic} interactions 
could not account for this increase of the $N=28$ gap. For instance,  it has 
been recently shown by Holt \etal~\cite{Holt10} that the $N=28$ gap is almost 
constant using two-body forces and grows when three-body forces are taken into 
account. A more recent work which uses interactions from chiral effective field theory, leads to the same conclusion:
the three-nucleon forces create a $N=28$
magic number~\cite{Hage12} and generate the shell closure in $^{48}$Ca.
More generally it was proposed by Zuker~\cite{Zuke03} that the 
implementation of a three-body force can solve several deficiencies in nuclear 
structure. 

\subsubsection{{\bf Neutron-neutron forces to create the $N=14$ sub-shell gap in O}}

The $N=14$ subshell gap, which is located between the $d_{5/2}$ and 
$s_{1/2}$ orbits,  also comes from the SO splitting and  
shares similar properties as the $N=28$ one. As shown in 
Figure~\ref{3body}, it grows by about $\delta G^{nn}(14)$=2.7~MeV from $^{17}$O to $^{23}$O 
as the neutron $d_{5/2}$ orbit is filled. Similar to the $N=28$ gap,  Equation~\ref{GAP3body} applies and
the large value of the $N=14$ gap at $N=14$ is due to the strongly 
attractive (repulsive) 
$V_{d_{5/2}~d_{5/2}}^{nn}$ ($V_{d_{5/2}~s_{1/2}}^{nn}$) monopoles terms.

\subsubsection{{\bf Could we predict the size of the $N=50$ shell gap in Ni?}}

From a phenomenological point of view,  we can reasonably expect that 
a similar increase in binding energy of major shells having $\ell$ and 
$s$ values aligned $\ell \uparrow$, i.e. $d_{5/2}$, $f_{7/2}$, $g_{9/2}$, 
$h_{11/2}$, will be occurring during their filling. This generally 
leads  to an increase of shell gap by: 

\begin{equation}\label{GAP3bodygeneral}
\delta G^{nn}(SO)=  x [V^{nn}_{\ell _\uparrow (\ell-2)_\uparrow}] - (x-1) [V^{nn}_{\ell _\uparrow \ell _\uparrow}] 
\end{equation}

In particular, we could look at what is expected for the $N=50$ shell gap 
from $^{68}$Ni to $^{78}$Ni, where fewer experimental data are available so 
far. Based on Equation~\ref{GAP3bodygeneral} the gap formed between the 
$g_{9/2}$ and $d_{5/2}$ orbits is expected to grow as the $1g_{9/2}$ orbit 
is filled. The relevant effective monopole terms 
($V_{g_{9/2}~g_{9/2}}^{nn} \simeq$ -200 keV 
and  $V_{g_{9/2}~d_{5/2}}^{nn} \simeq$+130 keV) can be extracted 
from the spectroscopy of $^{88,90}$Zr (see Figure~6 of Ref.~\cite{nous}). 
When renormalizing 
the monopole terms to $A \simeq 75$ (using the hypothesis that $V$ scales 
with $A^{-1/3}$) we find that the $N=50$ gap should increase by about 
3.1 MeV between $N=40$ and $N=50$.  The right-hand side of Figure~\ref{3body} 
displays the tentative change of the $N=50$ gap in the Ni isotopic chain. 
The $S_{n}$ value of $^{69}$Ni as well as the \emph{preliminary} energy of the $2d_{5/2}$ 
state at $\simeq$ 2.6 MeV  derived from Ref.~\cite{Mouk11} 
are used.  Note that a correction in binding energy is applied to $S_{n} (^{69}$Ni). Indeed it is likely that the $^{68}$Ni$_{40}$ ground state
has a mixed contribution with the  
$0^+_2$ state at 1.77 MeV. Taking  a mixing of $50\%(\nu g_{9/2})^2 + 50\%(\nu p_{1/2})^2$\footnote{This hypothesis comes from the
comparison with experimental results obtained for $^{90}_{40}$Zr, showing that the
configurations of its ground state and its $0^+_2$ state at 1.76 MeV are  
$50\%(\pi g_{9/2})^2 + 50\%(\pi p_{1/2})^2$.}, it implies 
that before mixing the $(p_{1/2})^2$ configuration of $^{68}$Ni is likely 
less bound by about 0.9~MeV. 
Thus starting from 2.6 MeV in $^{69}$Ni, the $N=50$ shell gap would 
reach (2.6+3.1=) 5.8 MeV in $^{78}$Ni. This is somehow larger 
than what extrapolated from the $N=50$ trend of binding energies~
\cite{nousPRC} and with the SM predictions of Ref.~ 
\cite{Siej12}. Note that  a small change in one of the $V^{nn}$ 
values  given above can drastically impact the value of the 
increased gap, as 10 neutrons ($x$ =10) are involved here between 
$N=40$ and $N=50$ (see Equation~\ref{GAP3bodygeneral}).

If this later increase of shell gap is confirmed, this will remarkably show 
that similar forces (including the three-body term) are at play in the 
various regions of the chart of nuclides to produce such large SO shell gaps. 
Even more \emph{a priori} surprising is the increase of all the $N=14$, 
$N=28$  and $N=50$ shell gaps by a \emph{similar} value of about 3~MeV. 
This comes from the fact that two competing terms are involved to enlarge the gap 
in Equation~\ref{GAP3bodygeneral}. The two monopoles decrease with 
$A^{-1/3}$ as the nucleus grows in size, but they are multiplied by the number of neutrons 
$x$ in the $\ell \uparrow$ orbit, which increases between $N=14$ 
($x=6$ in $d_{5/2}$), $N=28$ ($x=8$ in $f_{7/2}$), and $N=50$ 
($x=10$ in $g_{9/2}$). We could tentatively propose as an empirical rule that, 
during the filling of the $\ell \uparrow$ orbit, the variation of 
the major shell gaps of SO origin amounts to about 3 MeV.   Though rather crude, this  empirical rule could be used to predict the evolution
of the ($j\uparrow, (j-2)\uparrow$) orbits as a function of the filling of the $j\uparrow$ orbit. However, the 
pairing effect will dilute the orbital occupancy in heavier systems, 
as mentioned below for the $A=132$ region of mass.

\subsubsection{{\bf Evolution of the $N=82$ shell gap in Sn}}
 
As for the next shell $N=82$, we probably expect the 
same increase of the $h_{11/2} - f_{7/2}$ spacing while filling the 
$h_{11/2}$ orbit. However this increase of gap could hardly be evidenced 
experimentally as the $h_{11/2}$ orbit is located close to two  
low-$\ell$ orbits, $s_{1/2}$ and $d_{3/2}$. 
So the pairing correlations dilute the occupancy of the high-$j$ orbit 
with the two others, implying that the $h_{11/2}$ orbit is
gradually filling when adding 18 neutrons from $N=65$ to $N=82$.
Consequently the corresponding evolution of 
the $h_{11/2}$ binding energy is not only governed by the attractive 
$V_{h_{11/2}~h_{11/2}}^{nn}$ interaction, but also by the repulsive ones,
$V_{h_{11/2}~s_{1/2}}^{nn}$ and $V_{h_{11/2}~d_{3/2}}^{nn}$. 
Therefore, in this case, the SO empirical rule could 
be used as a guidance to constraint the spacing between the 
$\ell \uparrow$ and  $(\ell-2)\uparrow$ orbits  at the mean-field level, 
\emph{before} taking the role of correlations into account. 
It is important to point out that none of the self-consistent 
calculations reproduce the experimental location of the $\nu h_{11/2}$ 
orbit in the doubly magic $^{132}$Sn, i.e., when this orbit is completely
filled. All  mean-field models using Skyrme and Gogny forces or
relativistic NL3 and NL-Z2 forces predict that this
high-$j$ orbit is located just below the $N=82$ gap, the $\nu d_{3/2}$ 
orbit being more bound by about 1~MeV (see Figure~7 of Ref.~\cite{Bend03}), 
while the ground state of $^{131}$Sn has spin 3/2$^+$ and its first excited
state at 65~keV has spin 11/2$^-$.
One surmises that the lack of three-body forces is responsible for the 
discrepancy. Indeed these forces would strongly enhance the binding energy of 
the filled $\nu h_{11/2}$ orbit.

\subsubsection{{\bf A new sub-shell closure at $N=90$ ?}}

An interesting effect of the three-body forces would be the creation of 
a gap at $^{140}$Sn$_{90}$, between the $2f_{7/2}$ and $2p_{3/2}$ shells, 
as proposed by Sarkar \etal~\cite{Sarkar}. Indeed the neutron-neutron 
monopole terms (and possibly the three-body terms)  which intervene here 
are somehow similar to the ones at $N=28$. The only modifications are 
that the nodes of the wave functions differ, and that the monopole term 
for $A \simeq 48$ has to be downscaled to account for the reduction of 
the interaction energy as the nucleus grows in size at $A \simeq 140$. 
When using Equation~\ref{GAP3body}, the  suitable downscaled monopole terms, 
and the known $2f_{7/2} - 2p_{3/2}$ spacing of 0.85 MeV in 
$^{133}$Sn~\cite{Hoff96,Jone10} it is found that the $N=90$ gap is expected to grow by 
about 2.2 MeV while filling the $2f_{7/2}$ orbit. This leads to a subshell  
gap of about 3.1 MeV at $^{140}$Sn.  If established, this subshell gap 
would bring an additional credit to the mechanism of shell gap creation 
by three-body forces in another region of the chart of nuclides. In addition, 
this subshell could bring a local waiting point around $^{140}$Sn in 
the r-process nucleosynthesis.

In summary, the three-body forces are needed to create the neutron SO gaps. While
these gaps are not obtained when using microscopic effective interaction (see for
instance the discussion recently done by Smirnova \etal~\cite{Smir12}), the 
phenomenologically adjusted values of the TBME or the explicit implementation of
the three-body forces in SM approaches generate the expected neutron SO gaps.  

\subsection{Could we disentangle  tensor and spin-orbit 
forces in the $N=28$ region ?}\label{SOtensor}

\subsubsection{{\bf Introduction}}
The study of the nuclear structure around $N=28$ has revealed, using the shell 
model approach, that the tensor and two-body spin orbit interactions
are both at play to reduce the proton and neutron gaps far from stability. 
As proton orbits are degenerate at $N=28$, 
these two components could not be disentangled. To separate these components, 
we propose in this Section 
to study the evolution of $p$ SO splitting between $^{36}_{16}$S and  
$^{34}_{14}$Si (a bubble nucleus). We find a sizable reduction of the SO splitting there, which
is ascribed to the spin-orbit interaction (and not to tensor or central forces). This assumption 
comes from the fact that between these two nuclei 
protons are mainly removed from the  $2s_{1/2}$ ($\ell=0$) orbit, which does not exhibit a
specific orientation between the orbital momentum and the intrinsic spin value.

In a second part we propose to use the bubble nucleus $^{34}_{14}$Si to test 
the density and the isospin-dependent parts of the spin-orbit interaction in 
mean-field approaches which have never been tested so far.  
These properties of the spin-orbit interaction are of minor importance in 
the valley of stability but are crucial at the neutron drip-line and in 
superheavy nuclei.

\subsubsection{{\bf Studying the two-body spin-orbit interaction in the SM approach}}
  
\begin{figure}[t]
\begin{center}
\includegraphics[height=10cm]{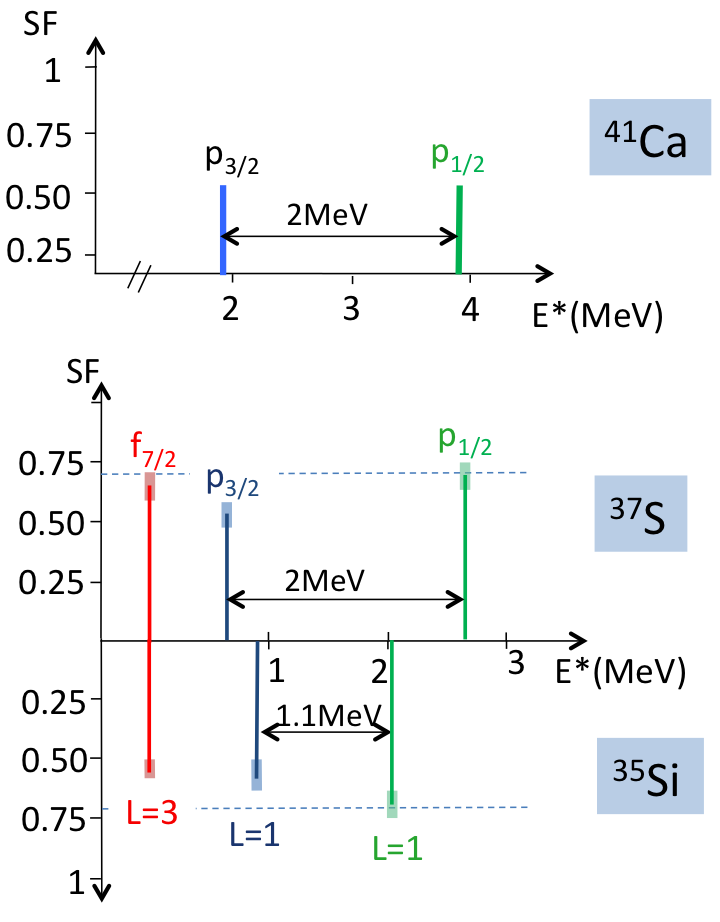}
\end{center}
\begin{center}
%\begin{minipage}{16.5cm}
\caption{Distribution of the major fragments of the single-particle strengths in 
the $N=21$ isotones. Spectroscopic factors
are represented in the y axis (adapted from Ref.~\cite{Jo11}).} 
\label{SOinCaSSi}
%\end{minipage}
\end{center}
\end{figure}

As  mentioned in Section~\ref{protonN28}, 
the proton $s_{1/2}$ and $d_{3/2}$ orbits are degenerate at $N=28$. 
The reduction of the $p$ SO splitting at $N=28$, between $^{49}$Ca and  $^{47}$Ar, is due to the combined 
effects of proton-neutron interactions involving several monopole terms 
having potentially tensor \emph{and} spin-orbit components. Applying Equation~\ref{SO} to the $p$ orbits, the SO splitting is:
%%%%%%%%%%%%%%%
\begin{equation} \label{SOp13}
\delta SO(p)= 1.33(V_{1d_{3/2} 2p_{1/2}}^{pn} - V_{1d_{3/2}
2p_{3/2}}^{pn}) + 0.66(V_{2s_{1/2} 2p_{1/2}}^{pn} -
V_{2s_{1/2} 2p_{3/2}}^{pn})
\end{equation}
%%%%%%%%%%%%%%%
The two contributions of this Equation can be estimated 
from the study of the evolution of the $p$ splitting in $N=21$ nuclei. As
 the  $s_{1/2}$  and $d_{3/2}$ proton orbits are 
separated by about 2.5 MeV (see $^{35}$P in Figure~\ref{s1d3inKClP}), one can therefore
assume that these proton orbits are filled \emph{sequentially} in the 
$N\simeq 20$ isotones. The pairing is providing a small dilution of 
occupancies among the two orbits as shown in Ref.~\cite{Khan85} and 
Figure~\ref{d5inP}. Between $^{41}_{20}$Ca$_{21}$ and $^{37}_{16}$S$_{21}$, 
four protons are removed.
The evolution of the $[\nu 2p_{3/2} -\nu 2p_{1/2}]$ SO splitting writes:
%%%%%%%%%%%%%%%
\begin{equation} \label{SO4}
\delta SO(p) \simeq 4 (V_{1d_{3/2} 2p_{1/2}}^{pn} - V_{1d_{3/2} 2p_{3/2}}^{pn})
\end{equation}
%%%%%%%%%%%%%%%
By looking first at the evolution of the  neutron $[2p_{3/2} - 2p_{1/2}]$ SO 
splitting between $^{41}$Ca$_{21}$ and $^{37}$S$_{21}$ in Figure~\ref{SOinCaSSi}, one investigates the 
role of proton-neutron interactions involving the $1d_{3/2}$ protons.  It is seen that the $p$ SO splitting amounts to 2 MeV in the two nuclei. 
It follows that the SO splitting does not change between 
$^{41}$Ca and $^{37}$S while 4 protons are removed from the $d_{3/2}$ shell, leading to
$\delta SO(p)\simeq 0$ and $ V_{1d_{3/2} 2p_{1/2}}^{pn} \simeq
V_{1d_{3/2} 2p_{3/2}}^{pn}$. We note that taking the full observed $p$ strength, the conclusion does not change. 
Indeed the SO splitting amounts to about 1.7 MeV in the two nuclei~\cite{Uozu94b,Eckl89}.
We conclude that the spin-dependent part of the $1d - 2p$ proton-neutron 
interaction is small. This is possibly due to the fact that the monopole terms  
$V_{1d_{3/2} 2p_{1/2}}^{pn} $ and $V_{1d_{3/2} 2p_{3/2}}^{pn}$ are weak in 
absolute value, as both their numbers of nodes and orbital momenta differ 
by one unit. 

From these observations, it was inferred also in Ref.~\cite{Gaud07} that  the change of $p$ SO 
splitting ($\delta SO(p)$) between $^{49}$Ca to 
$^{47}$Ar was ascribed to a modest depletion of the $s_{1/2}$ orbit  
by  about 0.66 protons (see also Section~\ref{neutron_espe}):
%%%%%%%%%%%%%%%
\begin{equation} \label{SOp3_N28}
\delta SO(p) \simeq   0.66 (V_{2s_{1/2} 2p_{1/2}}^{pn} -
V_{2s_{1/2} 2p_{3/2}}^{pn})
\end{equation} 
%%%%%%%%%%%%%%%
Thus at $N \simeq 28$ the effect of the two-body SO interaction is weak. 
The $N=21$ region is then more propicious to study this interaction as more 
protons are depleted in the $2s_{1/2}$ orbit  between $^{37}$S$_{21}$ and  $^{35}$Si$_{21}$. 

%%%%%%%%%%%%%%%%%%%%%%%%%%%%%%%%%%%%%%%%%%%%%%%%%%%%%%%%%%
%%%%%%%%%%%%%%%%%%%%%%%%%%%%%%%%%%%%%%%%%%%%%%%%%%%%%%
As for $^{35}$Si$_{21}$, energies and spectroscopic factors of 
its first $7/2^-$, $3/2^-$, $1/2^-$ and 
$5/2^-$ states were determined recently using the (d,p) transfer reaction 
in inverse kinematics at GANIL. They are compared to those of the isotone 
$^{37}$S$_{21}~$\cite{Jo11} in Figure~\ref{SOinCaSSi}. The observed $p$ SO splitting changes by about 900 keV between  
$^{37}$S and $^{35}$Si while protons are removed from the $2s_{1/2}$ orbit. 
This variation is observed in the \emph{major} fragment of the single-particle 
strength, which contains correlations beyond the monopole
terms. In order to unfold these correlations and to derive the change in monopole, the SM approach is used. 
While adjusting the monopole terms used in the SM calculations in order to reproduce 
the correlated experimental data~\cite{Nowa}, 
a decrease of the $p$ SO splitting from 2 MeV to 1.73 MeV is found for $^{37}$S and  
an increase  from 1.1 MeV to 1.35 MeV is found 
for $^{35}$Si. It follows that the change in 
$[p_{3/2} - p_{1/2}]$ spin-orbit splitting ($\delta SO(p)$) amounts to about 25\%, 
namely $(1.73-1.35)= 0.38$~MeV compared to the mean value $(1.73+1.35)/2=1.54$~MeV. 
This change 
is mainly ascribed to a change in the proton $2s_{1/2}$ 
occupancy which is calculated to be $\Delta (2s_{1/2})$=1.47 : 
%%%%%%%%%%%%%%%%%%%%%%%%%%%%%%%%%%%%%%%%%%%%%%%%%%%%%% 
\begin{equation} \label{SOp3}
\delta SO(p) \simeq   1.47 (V_{2s_{1/2} 2p_{1/2}}^{pn} -
V_{2s_{1/2} 2p_{3/2}}^{pn}) =~380~keV
\end{equation} 
%%%%%%%%%%%%%%%%%%%%%%%%%%%%%%%%%%%%%%%%%%%%%%%%%%%%%% 
Note that by using the monopole terms ($V_{2s_{1/2} 2p_{1/2}}^{pn}$ and 
$V_{2s_{1/2} 2p_{3/2}}^{pn})$ derived in Section~\ref{neutron_espe} and 
Figure~\ref{spe_N28}, one finds using Equation~\ref{SOp3} a consistent change in the $p$ SO 
splitting of 1.47(85+170)= 370 keV between $^{37}$S and $^{35}$Si\footnote{In practice the
monopole terms should be slightly renormalized according to the A$^{1/3}$ scaling rule 
between $A \simeq 48$ and $A \simeq 36$ region}. The present 
change in SO splitting is \emph{exclusively} due to the two-body spin-orbit force, 
as monopole terms involving an $\ell=0$ component, as it does here with the 
$s_{1/2}$ orbit, does not contain any tensor component~\cite{Otsu05}. The only worry for this extraction of two-body term 
comes from the fact that the energy 
and wave function of $1/2^-$ state in $^{35}$S, which is bound by about 0.5 MeV,  
may be influenced by the proximity of the continuum. 

\begin{figure}[t]
\begin{center}
\includegraphics[width=17cm]{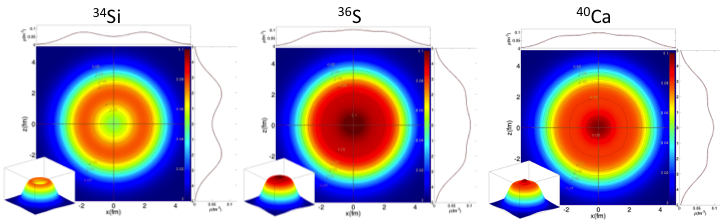}
\end{center}
\begin{center}
%\begin{minipage}{16.5cm}
\caption{Proton density profiles obtained with the RMF/DDME2 calculation 
in the $N=20$ isotones. From the left-hand picture,
it is seen that the $^{34}$Si nucleus displays a central depletion. 
By courtesy of J.P. Ebran.} 
\label{proton_densities}
%\end{minipage}
\end{center}
\end{figure}

Having an orbital momentum $\ell$=0, the $2s_{1/2}$ orbit is located in the 
center of the nucleus. With a weak $2s_{1/2}$ occupancy, $^{34}$Si is expected 
to have a depleted proton central density~\cite{Gras09} in the SM as well as in the MF approaches. This depletion is also
seen in the  density profile calculated using the RMF/DDME2 
interaction~\cite{Ebr12} in the left part of Figure~\ref{proton_densities}. 
%%%%%%%%%%%%%%%%%%%%%%%%%%%%%%%%%%%%%%%%%%%%%%%%%%%%%%%%%%
%%%%%%%%%%%%%%%%%%%%%%%%%%%%%%%%%%%%%%%%%%%%%%%%%%%%%%
On the other hand, the proton densities of  the other $N=20$ isotones
do not display any central depletion (see the middle and the right part of
Figure~\ref{proton_densities}). With this peculiarity
of density depletion \emph{and} isospin difference in central density (proton and neutron density profiles differ significantly), the  
$^{34}$Si nucleus is a good candidate to study the SO interaction in mean-field 
approaches, as described below.

\subsubsection{{\bf The spin-orbit interaction in the mean-field approaches}}

Interesting is to note that all theoretical approaches do not agree on the reason for the $N=28$ shell 
disappearance.  Li \etal~\cite{Li11} say that the '\emph{RMF model automatically reproduces the $N=28$ spherical shell gap 
because it naturally includes the spin-orbit interaction and the correct isospin dependence of this term'}. Therefore in the RMF approach '\emph{there is no need for a tensor interaction to reproduce the quenching of the spherical $N=28$ gap in 
the neutron-rich nucleiÉ'.}. The bubble nucleus $^{34}$Si will be used to ascertain these statements. 

Relativistic Mean Field (RMF) models introduced  description of the 
nuclear spin-orbit interaction~\cite{Ring96} in terms of mesonic 
degrees of freedom, treating the nucleons as Dirac particles. 
Using a non-relativistic reduction of the Dirac equation, the 
spin-orbit term writes:
%%%%%%%%%%%%%%%%%%%%%%%%%%%%%%%%%%%%%%%%%%%%%%%%%%%%%%
\begin{equation}\label{SORMF}
  V_\tau^{\ell s} (r) = - [W_1 \partial_r \rho_\tau (r) + W_2 \partial_r \rho_{\tau\neq \tau '} (r)] \overrightarrow{\ell} \cdot \overrightarrow{s}
\end{equation}
%%%%%%%%%%%%%%%%%%%%%%%%%%%%%%%%%%%%%%%%%%%%%%%%%%%%%%
in which $W_1$ and $W_2$ depend on the $\sigma, \omega, 
\rho$ meson coupling constants, and $\tau$ represents a proton or 
a neutron. Note that the $W_1$ and $W_2$ parameters have also density dependence in the 
RMF calculations that we neglect here. Thus beyond  the $\ell \cdot s$ term, the SO interaction 
contains a \emph{density}  dependence through $\partial_r \rho(r)$ and an \emph{isospin} dependence 
through the $W_{1}/W_{2}$ ratio. Both relativistic and non-relativistic 
mean-field (MF) approaches agree on a significant density dependence 
of the spin-orbit interaction.
However, while there is moderate isospin dependence of the SO 
interaction in the RMF approach ($W_1/W_2 \simeq$ 1), non-relativistic 
Hartree-Fock approaches have a large isospin dependence ($W_1/W_2$=2) 
provided by the exchange term~\cite{Sharma} of the nuclear interaction. 
Beyond the fundamental importance to describe the spin orbit interaction, 
the density and isospin dependences of the spin-orbit interaction have important 
consequences to model nuclei close to the drip-line~\cite{Lala98}, to predict the isotopic 
shifts in the Pb region~\cite{Rein95}
and to  describe superheavy nuclei which display a central-density depletion~\cite{Bend99}.

Having a significant proton-density depletion and a proton-neutron 
density asymmetry (there is no neutron central depletion), the $^{34}$Si nucleus is a good candidate to study both 
the density and isospin dependence of the spin-orbit interaction as 
proposed in Ref.~\cite{Jo11}. Indeed RMF models give a large change 
in SO splitting by about 80\% between $^{37}$S and $^{35}$Si while MF 
models find a more modest change of about 30\% for 
$\Delta (2s_{1/2})$ =1.5~\cite{Jo11}. 

The presently observed  decrease of the SO splitting between $^{37}$S 
and $^{35}$Si by about 25\% indicates that there is a density 
dependence of the SO interaction, otherwise no change in SO 
splitting would have been found. This modest change  
is in better agreement with theoretical models  having a \emph{large} isospin 
dependence, such as most of the MF models. Though extremely preliminary, this 
discovery would suggest that the treatment of the isospin dependence of the SO interaction 
in the RMF approach is not appropriate. 
If true this would have important consequences to model nuclei which 
are sensitive to the isospin dependence of the SO interaction. We give here two examples.

In some superheavy nuclei, large proton and neutron central-density depletion may be 
present. They originate from the coulomb repulsion between protons
and from the filling of high-$j$ neutron orbits which are located at the nuclear surface 
only~\cite{Bend99}. Assuming no isospin dependence of the SO interaction as in the RMF 
approach, the proton and neutron central-density depletions \emph{mutually} reduce the proton SO 
splitting for ($\ell=$1, 2, and 3) orbits which are probing  the interior of the 
nucleus~\cite{Bend99}. In this case, the $Z=114$ gap, formed between the proton 
$f_{5/2}$ and $f_{7/2}$ would be 
significantly reduced to the benefit of a large shell gap at $Z=120,N=172$. 
Conversely, a rather large isospin dependence of the SO interaction would not 
lead to strong shell gaps~\cite{Bend01} but rather to sub-shells the rigidity of 
which is further eroded by correlations. 
The second example deals upon nuclei close to the neutron drip-line. As they are 
expected to have a smoothened neutron density at their surface, the spin-orbit 
of high-$j$ orbits is expected to be weakened. 

In the RMF approach the SO interaction is 
weaker by about up to 65\% as compared to MF approaches when reaching the drip -line~\cite{Lala98}. 
This change between RMF and MF models seems to be due to 
the different treatment of the isospin dependence of the SO interaction~\cite{Lala98}.  
This change in SO interaction could have important consequences to model the evolution 
of the  $N=28$ shell closure far from stability as well as other SO magic numbers  
along which the r-process nucleosynthesis occurs~\cite{Pfei01}. We note here that other competing 
effects come into play to modify shell structure when approaching the drip-line, 
as mentioned in the following Sections. 

\subsection{Would other SO magic numbers 14, 50, 82  be 
disappearing far from stability as $N=28$ does and if not 
why ?}\label{otherSO}
 
\subsubsection{{\bf Introduction}}

In this Section we look whether the SO magic gaps $14,50,82$ are progressively vanishing
when moving far from stability in the same manner as $N=28$ does toward $^{42}$Si. We start with the analogy of the  evolution of the $N=14$ shell gaps between  $^{22}$O and $^{20}$C and pursue with the evolution of the SO magic number in the $^{132}$Sn nucleus at $N=82$. We show that, surprisingly, while the same forces are involved in all regions, the $N=82$ gap remains very large contrary to the $N=14$ and $N=28$ ones. The $^{78}$Ni nucleus, which lies in between the $^{42}$Si and $^{132}$Sn nuclei, is briefly discussed as well. 

\subsubsection{{\bf Disappearence of the $N=14$ and $N=28$ gaps far from stability}}

\begin{figure}[t]
\begin{center}
\includegraphics[height=12cm]{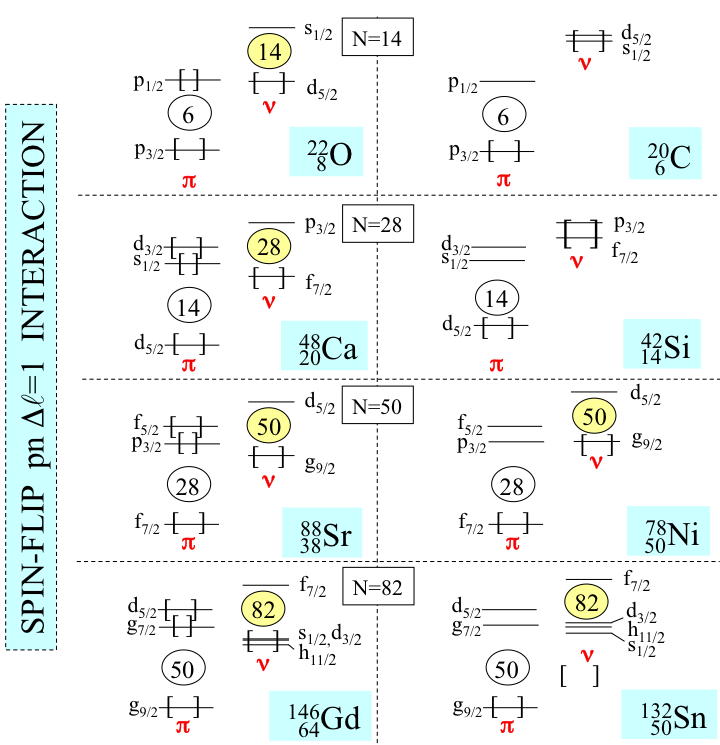}
\end{center}
\begin{center}
%\begin{minipage}{16.5cm}
\caption{Schematic evolution of the SO shell gaps at 
$N=$14, 28, 50 and 82  as a function of the filling 
of the proton $\ell \downarrow$ orbits, with 
$\ell =$1, 2, 3 and 4, respectively.  
While the $N=14$ and $N=28$ magic gaps vanish in nuclei far from the
stability valley (right column), the $N=50$ gap slightly decreases and 
the $N=82$ one remains very robust.}  
\label{SOgeneral}
%\end{minipage}
\end{center}
\end{figure}

The $N=28$ gap vanishes progressively by the removal of 6 protons 
from the doubly magic $^{48}$Ca to the deformed $^{42}$Si nucleus. 
As far as tensor forces are concerned, their action cancels 
in a spin saturated valence space, i.e., when orbits with 
aligned and anti-aligned spin orientations with respect 
to a given angular momentum are filled. In the present 
valence space for instance, the filling of the proton  
$f_{7/2}$ and $f_{5/2}$  annihilates the tensor effects 
on the filled $d_{5/2}$ and $d_{3/2}$ orbits. As the $^{42}$Si nucleus 
is a spin unsaturated system in protons and neutrons, tensor 
effects are maximized, leading to reductions of both the 
proton $d_{3/2} - d_{5/2}$ and neutron $f_{7/2} - f_{5/2}$ splittings.  
Similar spin unsaturated nuclei exist in nature, such as 
(i) $^{20}$C ($\pi p_\uparrow$ and $\nu d_\uparrow$ filled but 
not  $\pi p_\downarrow$ and $\nu d_\downarrow$),  (ii) $^{78}$Ni  
($\pi f_\uparrow$ and $\nu g_\uparrow$ filled but not  
$\pi f_\downarrow$ and $\nu g_\downarrow$) and (ii) $^{132}$Sn 
($\pi g_\uparrow$ and $\nu h_\uparrow$ filled but not  
$\pi g_\downarrow$ and $\nu h_\downarrow$). These examples are 
illustrated in Figure~\ref{SOgeneral}. 
We would  expect that the behavior of all these four mirror valence nuclei 
is similar, but it is not the case.
%%%%%%%%%%%%%%%%%%%%%%%%%%%%%%%%%%%%%%%%%%%%%%%%%%%%%%%%%%
%%%%%%%%%%%%%%%%%%%%%%%%%%%%%%%%%%%%%%%%%%%%%%%%%%%%%%

Starting with the first example, $N=14$, it was found in Ref.~\cite{Stan08} 
that the energy of the first 2$^+$ state drops by a factor of two  
between $^{22}$O (3.2 MeV~\cite{Stan04}) and $^{20}$C (1.588 MeV). 
Then the reduction of the $N=14$ gap, formed between the neutron $d_{5/2}$ 
and  $s_{1/2}$ orbits, could be ascribed to the two monopole interactions 
involved when removing 2 protons from the $p_{1/2}$ orbit between 
$^{22}$O and $^{20}$C, i.e.:
%%%%%%%%%%%%%%%
\begin{equation}\label{GAP14}
\noindent \delta G^{pn}(14) \simeq 2(V_{1p_{1/2} 2s_{1/2}}^{pn} - V_{1p_{1/2} 1d_{5/2}}^{pn}) 
\end{equation}
%%%%%%%%%%%%%%%
We can estimate the reduction of the $N=14$ gap by looking at similar 
forces in the valley of stability. The ordering of the $d_{5/2}$ and  
$s_{1/2}$ is exchanged  between $^{17}$O (g.s. $5/2^+$, $1/2^+$ at 
870 keV) and $^{15}$C (g.s. $1/2^+$, $5/2^+$ at 740 keV). Such a  
swapping of about 1.6 MeV between the two neutron orbits means that 
$| V_{1p_{1/2} 1d_{5/2}}^{pn} | >> | V_{1p_{1/2} 2s_{1/2}}^{pn} | $.
The large value of the $| V_{1p_{1/2} 1d_{5/2}}^{pn} |$ monopole can 
be ascribed 
to the attractive part of the tensor term.
Assuming that the two monopoles are of similar intensity between 
$A=16$ and $A=22$, the $N=14$ gap is expected to be reduced by 1.6 MeV between 
$^{22}$O and $^{20}$C, starting from a value of about 4 MeV in $^{22}$O 
(see Figure~\ref{3body}). This large reduction of the $N=14$ spherical 
gap, by the removal of only two protons,  can easily account  for 
the decrease in 2$^+$ energy in $^{20}$C. 

For the $N=28$ shell gap, the removal of 6 protons is required to 
change drastically the nuclear structure. The  effect of residual 
forces per nucleon is less important than in the 
lighter nuclei. It is also noticeable that the size of the $N=28$ 
gap (4.8 MeV) is about 1 MeV larger than that of $N=14$. These two 
effects, weaker monopole terms and increasingly larger SO splitting, 
could explain why the vanishing of the shell gap occurs further from stability at $N=28$.

\subsubsection{{\bf Persistence of the $N=82$ gap in $^{132}$Sn}}

On the other hand, $^{132}$Sn has all the properties 
of a doubly-magic nucleus with a high $2^+$ energy of 4.04 MeV~\cite{Foge94}, due to 
a large shell gap~\cite{nous}. Large 
spectroscopic factors are found in the $^{132}$Sn (d,p) $^{133}$Sn 
reaction~\cite{Jone10,Jones_contrib}, which is a further confirmation that $^{132}$Sn 
is a closed-shell nucleus. Note that the $N=82$ gap is not strictly 
formed between the neutron $h_{11/2}$ and $f_{7/2}$ orbits, as the 
$h_{11/2}$ orbit lies below the $d_{3/2}$ one for $Z=50$~\cite{Foge04}. 
This arises from the fact that the $h_{11/2}- h_{9/2}$ SO splitting 
is larger than the other SO splittings 
$g_{9/2}- g_{7/2}$, $f_{7/2}- f_{5/2}$ and $d_{5/2}- d_{3/2}$ in 
the $N=50$, $N=28$ and $N=14$ regions respectively. This increasingly 
large splitting with $\ell$ value is due to the $\overrightarrow{\ell} \cdot \overrightarrow{s}$ term 
of Equation~\ref{SORMF}. 

\subsubsection{{\bf The $N=50$ shell gap in $^{78}$Ni}}

In the $N=50$ case, the rigidity of the $^{78}$Ni nucleus with respect to 
quadrupole excitations is not known so far. Based on the study of 
spectroscopic information between $Z=30$ and $Z=38$, the evolution 
of the $N=50$ shell gap has been studied in Ref.~\cite{nousPRC}. A 
rather modest reduction of  about 0.55 MeV was proposed between $Z=38$ and
$Z=28$. Nevertheless, as this reduction is combined with that of the $Z=28$ 
gap~\cite{Siej12}, it is not absolutely sure that $^{78}$Ni 
would remain spherical in its ground state.  

\subsubsection{{\bf Conclusions}}
Though similar tensor and spin-orbit forces are at play for 
$N=14$, 28, 50 and 82, the corresponding shell closures do not have 
the same behavior when going far from stability. Quadrupole 
correlations dominate in  $N=14$ and $N=28$ over the 
spherical shell gaps, while  the $N=82$ gap remains remarkably rigid 
in $^{132}$Sn. This feature likely comes from two effects : (i) the SO 
shell gaps become larger for higher $\ell$ orbits and (ii) the monopole 
terms involved to reduce the shell gap are weakened in heavy nuclei 
because of the larger sizes of the orbits.  In the next Section we explore the 
forces which come into play even further from stability.

\subsection{Which forces come into play in nuclei located below 
$^{42}$Si,  $^{78}$Ni and $^{132}$Sn?}\label{below42Si}

\subsubsection{{\bf Introduction}}
Advances in radioactive beam productions offer the possibility to produce 
and study nuclei at the limit of particle stability. 
There, new forces are involved and the interaction with continuum states 
should progressively play a decisive role. We start this Section by 
a qualitative analysis of the nuclear forces involved at $N=28$ below 
$^{42}$Si. We propose that this gap is further reduced, and illustrate our 
purpose by looking at recent experimental studies which evidence the 
inversion between the $f_{7/2}$ and $p_{3/2}$ orbits. We generalize this 
mechanism to other shell gaps, $N=50$ and $N=82$, and propose astrophysical 
consequences for the r-process nucleosynthesis below $^{132}$Sn.

\subsubsection{{\bf The $N=28$ shell gap below $^{42}$Si}}

It was shown in Refs.~\cite{Tara87,Baum07} that the neutron drip line extends 
further $N=28$ in the Si and Al isotopic chains, i.e., at least up to 
$^{44}$Si$_{30}$ and $^{43}$Al$_{30}$, respectively.
If the $N=28$ shell gap 
was large, a sudden drop in $S_{2n}$ would be found there and the 
drip-line would lie at $N=28$ and not beyond.  To give an example,
the neutron drip line is exactly located at a magic shell in 
the O chain, where $^{24}$O is bound by about 3.6 MeV and 
$^{25}$O is unbound by about 770 keV \cite{Hoff08}.  So in this 
case, $^{24}$O is a doubly-magic
nucleus~\cite{Kanu02,Hoff09,Tsho12} and the spherical shell gap is 
large enough to hamper the onset of quadrupole correlations. 
By contrast the fact that the drip line extends further from stability 
at $N=28$ is an indication for the presence of deformed nuclei which 
gain in binding energy due to correlations. This also hints for a 
further reduction of the $N=28$ shell gap. This hypothesis is hard 
to prove in a direct way as it would require to determine the size of the 
spherical shell gap there, which is out of reach so far, and which would 
not be an observable when nuclei are deformed. Therefore we shall 
use qualitative arguments as well as an example taken from another 
region of the chart of nuclides where similar forces are present 
to predict the behavior of the $N=28$ shell gap further from stability. 
With this in mind, we will propose some extrapolations to other regions 
of the chart of nuclides.

\begin{figure}[t]
\begin{center}
\includegraphics[height=10cm]{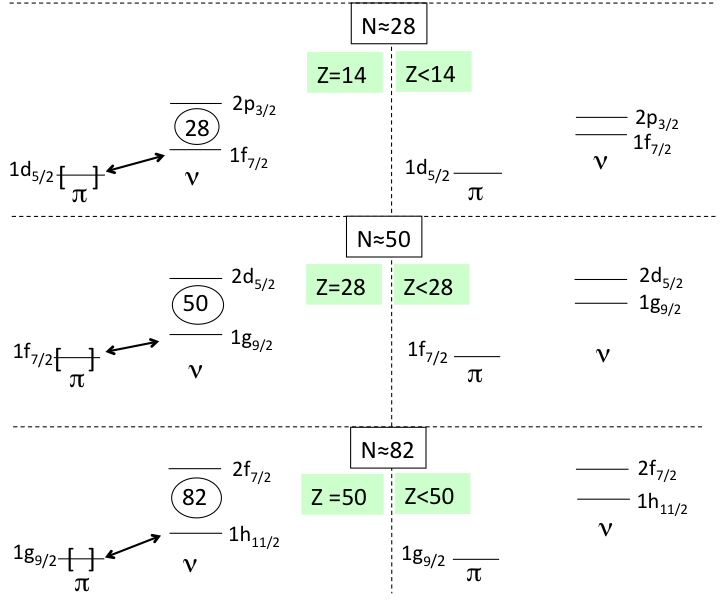}
\end{center}
\begin{center}
%\begin{minipage}{16.5cm}
\caption{Schematic evolution of the SO shell gaps at 28, 50 and 
82, as a function of the filling of the proton 
$\ell\uparrow$ orbits, with $\ell =$2, 3 and 4, respectively.  
The attractive pn interaction $1d_{5/2} - 1f_{7/2}$ is larger 
than the $1d_{5/2} - 2p_{3/2}$ one. Therefore the $N=28$ gap is 
larger when the $d_{5/2}$ orbit is completely filled (left part). Conversely, 
when emptying the $\pi d_{5/2}$ orbit (i.e., going very far from stability) 
the $N=28$ gap is expected to vanish (right part). The same mechanism is foreseen 
for the $N=$50 (82) gap when the $\pi f_{7/2}$ ($g_{9/2}$) orbit is empty.}  
\label{npforces_below42Si}
%\end{minipage}
\end{center}
\end{figure}

Below $^{42}$Si the change of the $N=28$ shell gap between the $1f_{7/2}$ 
and $2p_{3/2}$ orbits is driven by the difference between the 
$V_{1d_{5/2} 2p_{3/2}}^{pn}$ and $V_{1d_{5/2} 1f_{7/2}}^{pn}$ 
monopole terms. On top of this monopole-driven effect, correlations 
play an important role in these nuclei.  For these two monopole 
terms, the proton and neutron spin orientations are aligned with 
the orbital momentum, and their difference in angular momentum is 
one unit of $\hbar$ (then the main part of the proton-neutron interaction 
comes from the central term of the nuclear force). On the other hand, the numbers of 
nodes in their wave functions are different. The fact that  the $1f_{7/2}$ and 
$1d_{5/2}$ wave functions have the same number of nodes leads to 
a larger radial overlap of the wave functions, and a larger monopole 
term $V_{1d_{5/2} 1f_{7/2}}^{pn}$ as compared to  
$V_{1d_{5/2} 2p_{3/2}}^{pn}$. Consequently when the $1d_{5/2}$ orbit is 
completely filled, the $1f_{7/2}$ orbit has a larger binding energy than 
that of the $2p_{3/2}$ one. Conversely, when removing the six protons from 
the $1d_{5/2}$ orbit, the neutron $N=28$ gap further shrinks. Note that the
calculations of Ref.~\cite{Smir10} predict a decrease by 1.6 MeV down to
$^{36}$O$_{28}$. Of course this nuclei  lies beyond the drip-line, but this gives 
the monopole trend for the evolution of the $N=28$ gap below $^{42}$Si 
(see Figure~\ref{npforces_below42Si}). 
%%%%%%%%%%%%%%%%%%%%%%%%%%%%%%%%%%%%%%%%%%%%%%%%%%%%%%%%%%
%%%%%%%%%%%%%%%%%%%%%%%%%%%%%%%%%%%%%%%%%%%%%%%%%%%%%%

In addition to this monopole effect, the nuclear-potential well will 
become more diffuse when moving towards the drip-line. It was proposed by 
Dobaczweski \etal~\cite{Doba94} that this would bring the low $\ell$ 
orbits more bound as compared to the high $\ell$ ones, which 
could be viewed schematically as a reduced $\ell^2$ term in the 
Nilsson potential. Then, an inversion 
of the two negative-parity orbits ($f$ and $p$) could also happen, 
as suggested by several experimental works described below.

While the $\nu 1f_{7/2}$ orbit lies below the $\nu 2p_{3/2}$ orbit in most of 
the nuclei in the chart of nuclides (giving rise to the $N=28$ shell gap), 
their ordering is reversed as soon as the $\pi d_{5/2}$ shell starts to empty,
i.e., for $Z < 14$. The feature is well documented in the $N=15$ and $N=17$ 
isotones, not far from the stability valley. Indeed, the first negative-parity state of
$^{29,31}_{14}$Si$_{15,17}$ has $I^\pi=7/2^-$ while the 3/2$^-$ state is located
above. On the other hand, the situation is reversed in $_{12}$Mg and $_{10}$Ne 
isotopes, as described now.   
In the $^{25}$Ne$_{15}$ nucleus~\cite{Catf10}, the $2p_{3/2}$ orbit lies 
about 1 MeV \emph{below} the $1f_{7/2}$ orbit. This was determined 
by using the $^{24}$Ne(d,p)$^{25}$Ne transfer reaction in inverse 
kinematics at the GANIL facility. Similar results were recently 
obtained at GANIL in $^{27}$Ne$_{17}$,  its first excited state at 765~keV
having $I^\pi=3/2^-$~\cite{Brow12}, the $7/2^-$ state being about 1 MeV above. 
Note that the assignment of the 
$3/2^-$ state was first proposed in Refs.~\cite{Obertelli, Terry}. 
In the neighboring isotone $^{29}$Mg, the second excited
state at 1095~keV was observed from the study of the $\beta$ decay of  $^{29}$Na. It
could not be explained in the framework of shell-model calculations using
the $sd$ space, and consequently it was assigned as a $I^\pi=3/2^-$ intruder 
state~\cite{Baum87}.
To our knowledge, two other 
experimental works are suggesting this inversion between the 
$1f$ and $2p$ orbits. By using Coulomb break-up technique 
at RIKEN, Nakamura \etal~\cite{Naka10} found that the ground-state 
wave function of $^{31}$Ne was dominated by an $\ell=1$ component. 
Minomo \etal~\cite{Mino12} propose a contribution of a $p_{3/2}$ neutron 
using antisymmetrized molecular dynamics calculations. 
Moreover Wimmer \etal argued that excitations to the 
$2p_{3/2}$ (rather than the $1f_{7/2}$) are required to account 
for the configuration of the $0^+_{1,2}$ states in $^{32}$Mg 
determined through two-neutron transfer reaction at 
CERN/ISOLDE~\cite{Wimm10}. 
In these  examples, in which the proton $1d_{5/2}$ orbit is not yet 
fully filled, the $N=28$ gap does not exist and the $\nu 2p_{3/2}$ orbit 
even lies below the $\nu 1f_{7/2}$ orbits. 
Conversely, this gives a further indication that the proton 
$1d_{5/2}$ has to be filled to bind the $\nu 1f_{7/2}$ orbit enough
to create a $N=28$ gap which is present in heavier nuclei. 
Note that, from a theoretical point of 
view, this inversion was also proposed by Utsuno~\etal in 
1999~\cite{Utsu99}. 

\subsubsection{{\bf Extrapolations to other shells $N=50$ and $N=82$}}

We can reasonably assume that the present mechanism for the 
reduction of the $N=28$ gap is robust, as it comes from the radial 
part of the wave functions and from the effect of a more diffused 
potential well at the drip-line. As far as the first effect is concerned, it is also observed
from the properties of the bare proton-neutron forces \cite{Morten}.
We therefore also expect the reduction  of the $N=50$ gap 
below $^{78}$Ni and the reduction of the $N=82$ gap below $^{132}$Sn, 
as protons are removed from the $1f_{7/2}$ and $1g_{9/2}$ orbits, 
respectively (see Figure \ref{npforces_below42Si}). While the $N=82$ gap is still present in the $^{132}$Sn 
nucleus, it will progressively be reduced further from stability as 
ten protons are removed from the $1g_{9/2}$ orbit to reach $^{122}$Zr. 
The onset of quadrupole correlations across the $N=82$ shell gap 
would occur if the neutron gap is reduced enough. This would bring a 
smoothening in the $S_{2n}$ trend, as observed in Figure~\ref{S_2n} 
for the $N=28$ isotones around $^{44}$S. This has potentially important 
consequences on the location of the $(n,\gamma)-(\gamma,n)$ 
equilibrium in the rapid neutron-capture  process (r-process). 
This equilibrium occurs when neutron captures compete with 
photo-disintegrations. At this point, further neutron captures 
cannot occur as nuclei are immediately destroyed. Therefore the 
explosive process is stalled for a moment at such waiting-point 
until $\beta$-decay occurs. When the drop in $S_{2n}$ value is 
abrupt (for a large shell gap), the waiting-point nuclei are found 
\emph{at} the closed shell.  Then these nuclei are the main genitors of 
stable r-elements when decaying back to stability. When the shell 
gap is reduced, the location of r-progenitors is more extended and 
shifted to lower masses in a given $Z$ chain, changing the fit of 
the r-process abundance curve accordingly (see for instance 
Ref.~\cite{Pfei01}). 

To summarize this Section, we propose that both from the nuclear two-body forces and from
the mean-field point of views a further reduction of the $N=28$ shell gap will be occurring 
below $^{42}$Si. By analogy, a reduction of the $N=50$ and $N=82$ shell gaps is anticipated, 
with possible important consequences for the r-process nucleosynthesis in the later case.

\subsection{Are proton-neutron forces significantly modified when 
approaching the drip line ?} 

\subsubsection{{\bf Introduction}}

In the previous Sections we assumed that the two-body proton-neutron forces 
(or the monopole terms) remain of similar intensity when reaching 
the drip line. The SM description does not allow a self-consistent change of the monopole terms as a
function of binding energy. However the assumption of fixed monopole terms in a wide valence space 
is at best crude and probably wrong. 
Indeed at the  neutron drip-line, valence protons are very deeply bound 
and their wave functions are well confined. On the other hand, 
neutrons, which are close to be unbound, have more dilute wave 
functions and interactions with continuum states occur~\cite{Doba07}. It follows that the
radial part of the wave functions of the valence proton and neutron have weaker overlap, 
leading to a reduced effective interaction. 
One could question by how much this interaction is reduced ?

The $_{9}$F isotopic chain is a target choice to study  changes in proton-neutron forces 
when approaching the drip line. 
Indeed the drip-line occurs at $N=16$ in the $_{8}$O chain, i.e., at $^{24}$O, while 
the $^{31}$F$_{22}$ nucleus is still bound. Thus the addition of a \emph{single}
proton allows to bind up to 6 neutrons more. It follows that, though 
occurring between nuclei showing large asymmetry in neutron and proton binding energy, 
the corresponding proton-neutron forces should be large enough to bind 
$^{31}$F$_{22}$\footnote{Beyond proton-neutron interactions, we remind here 
the important role of neutron-neutron forces as well.}. Up to $^{29}$F$_{20}$, 
the same proton-neutron $d_{5/2}-d_{3/2}$ is expected to intervene. This strong 
monopole force is well known to produce significant changes closer to stability 
in the shell evolution of the $N=16$ and $N=20$ gaps. How is this force modified 
far from stability ? The recent determinations of the binding energies of the 
(unbound) $^{28}$F$_{19}$ ~\cite{Chri12} and the bound $^{29}$F$_{20}$~\cite{ref_laurent} 
offer the possibility to start answering these questions. However error bars on 
binding energies are still a bit too large at present to draw firm conclusions.

In this Section we propose to study the spectroscopy of the $^{26}_{9}$F$_{17}$ to study the proton-neutron $d_{5/2}-d_{3/2}$ interaction when approaching the drip line. We first explain the motivation of studying this nucleus and subsequently use the new  spectroscopic information required to study the proton-neutron force. 

\subsubsection{{\bf Experimental study of $^{26}_{9}$F}}

\begin{figure}[t]
\begin{center}
\includegraphics[width=16cm]{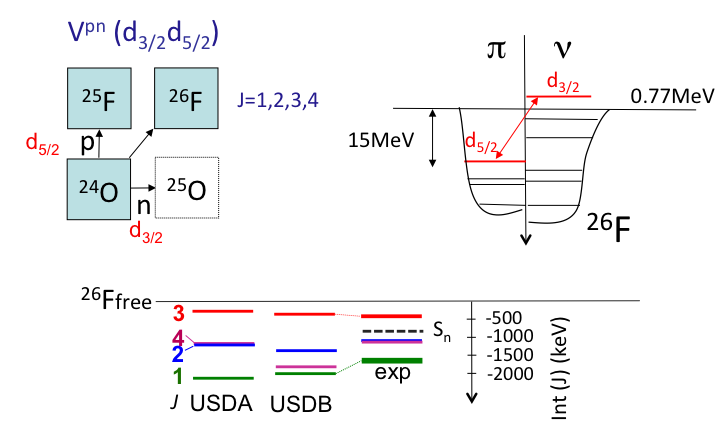}
\end{center}
\begin{center}
%\begin{minipage}{16.5cm}
\caption{Schematic view of the low-energy configurations in 
$^{26}$F, which could be approximated as a core of $^{24}$O on 
top of which a deeply bound proton $d_{5/2}$ and an unbound 
neutron $d_{3/2}$ are added. This could bring, after having 
treated the role of correlations, information on the proton-neutron 
$d_{5/2}- d_{3/2}$ interaction close to the drip-line. In the 
bottom part, values of int(J) extracted from the SM
calculations are compared to the presently known experimental 
values for $J^\pi=1^+-4^+$.}
\label{npforces_26F}
%\end{minipage}
\end{center}
\end{figure}

The $^{26}_{9}$F$_{17}$ nucleus is a suitable choice for studying the proton-neutron $d_{5/2}-d_{3/2}$ interaction
for at least three arguments: (i) it is bound by only 0.80(12)~MeV,  (ii) as the first excited state in 
$^{24}$O lies at 4.47~MeV~\cite{Hoff09}, $^{26}$F  can be viewed 
as a closed $^{24}$O core plus a \emph{deeply bound} proton in 
the $d_{5/2}$ orbit ($S_p$($^{25}$F)$\simeq$ -15~MeV) and an 
\emph{unbound} neutron in the $d_{3/2}$ orbit 
($S_n$($^{25}$O)$\simeq$ +0.77~MeV~\cite{Hoff08}), (iii)  the $\pi d_{5/2}$ and 
$\nu d_{3/2}$ orbits are rather well separated from other orbits 
which limit correlations of pairing and quadrupole origin. Binding 
energies BE($^{26}$F)$_J$  of the \emph{full} multiplet of $J=1-4^+$ 
states arising from the $\pi d_{5/2} \otimes \nu d_{3/2}$ coupling 
in $^{26}_{9}$F$_{17}$ are needed to determine the role of the 
coupling to the continuum on the mean $\pi d_{5/2} \nu d_{3/2}$ 
interaction energy ($int$), defined in Equation~(\ref{int}), as well as 
on the residual interaction which lifts the degeneracy between  
the components of the multiplet $int(J)$ defined in Equation~(\ref{IntJ}). 
The mean interaction energy ($int$) writes:
%%%%%%%%%%%%%%%%%%%%%%%%%%%%%%
\begin{equation}\label{int}
int= \sum\frac{(2J+1)\times int(J)}{(2J+1)}
\end{equation}
%%%%%%%%%%%%%%%%%%%%%%%%%%%%%%
In this Equation $int(J)$ term expresses the difference between the experimental 
binding energy of a state $J$ in $^{26}$F (BE($^{26}$F)$_J$) and 
that of the $^{24}$O+1p+1n system, $BE(^{26}F_{free})$, in which the valence 
proton and neutron do not interact. It writes:
%%%%%%%%%%%%%%%%%%%%%%%%%%%%%%
\begin{equation}\label{IntJ}
int(J)= BE(^{26}F)_J - BE(^{26}F_{free}),
\end{equation}
%%%%%%%%%%%%%%%%%%%%%%%%%%%%%%
where 
%%%%%%%%%%%%%%%%%%%%%%%%%%%%%%
\begin{equation}
BE(^{26}F_{free}) = BE(^{24}O)_{0^+} + BE(\pi d_{5/2}) + BE(\nu d_{3/2})
\end{equation}
%%%%%%%%%%%%%%%%%%%%%%%%%%%%%%
Assuming that 
%%%%%%%%%%%%%%%%%%%%%%%%%%%%%%
\begin{equation}
BE(\pi d_{5/2})= BE(^{25}F)_{5/2^+}-BE(^{24}O)_{0^+}
\end{equation}
and
\begin{equation}
BE(\nu d_{3/2})= BE(^{25}O)_{3/2^+}-BE(^{24}O)_{0^+}
\end{equation}
%%%%%%%%%%%%%%%%%%%%%%%%%%%%%%
we obtain:
%%%%%%%%%%%%%%%%%%%%%%%%%%%%%%
\begin{equation}
BE(^{26}F_{free}) = BE(^{25}F)_{5/2^+}+BE(^{25}O)_{3/2^+}-BE(^{24}O)_{0^+}
\end{equation}
%%%%%%%%%%%%%%%%%%%%%%%%%%%%%%
Following the particle-particle coupling rule, the values of 
$int(J)$ should display a parabola as a function of $J$ in which 
$| int(1) | $ and $ | int(4) |$ are the largest, as proton and 
neutron maximize their wave-function overlap for these two values of 
total angular momentum, leading to the largest attractive interactions.  
The values of $int(J)$ obtained from shell model calculations 
(see bottom part of Figure \ref{npforces_26F}) indeed form a parabola, 
the $|int(3)|$ value being the lowest.

%%%%%%%%%%%%%%%%%%%%%%%%%%%%%%%%%%%%%%%%%%%%%%%%%%%%%%%%%%
%%%%%%%%%%%%%%%%%%%%%%%%%%%%%%%%%%%%%%%%%%%%%%%%%%%%%%
The atomic masses of $^{26,25}$F and $^{24}$O were determined by 
Jurado \etal \cite{Jura07}. They were used to determine the binding 
energies BE values used in the previous Equations. The  
$BE(^{25}O)_{3/2^+}-BE(^{24}O)_{0^+}$=+0.77~MeV difference is derived
from the work of Hoffman \etal~\cite{Hoff08}. The spin of the ground 
state of $^{26}$F was determined to be $I^\pi=1^+$ by Reed \etal ~\cite{Reed99} from the
fact that $I^\pi=0,2^+$ states were populated in its $\beta$ decay. The 
first-excited state of $^{26}$F has been recently identified at 
657(7)~keV excitation energy by Stanoiu \etal~\cite{Stan12} at GANIL using  
in-beam $\gamma$-ray detection
combined to the double fragmentation method. Moreover, an unbound state 
lying 270~keV above the neutron emission threshold has been 
proposed by Franck  \etal following
a nucleon-exchange reaction at NSCL~\cite{Fran11}. This state 
is likely to be the $I^\pi=3^+$ state. 
Finally a long-lived $J=4^+$ isomer, $T_{1/2}=2.50(5)$ms, has 
been discovered at 642~keV by Lepailleur \etal at GANIL \cite{Lepailleur}. It decays by an $M3$ internal 
transition to the $J=1^+$ ground state (86\%), and by beta-decay 
to $^{26}$Ne or beta-delayed neutron emission to $^{25}$Ne. 
Gathering the measured binding energies of the $J=1-4^+$ multiplet 
in $^{26}_{9}$F in Figure~\ref{npforces_26F} \footnote{The atomic mass 
of  $^{26}$F measured in~\cite{Jura07} was shifted to a larger binding 
energy (downward) to take into account the probable contamination 
of 40\% in isomeric state.} we find that the components of the multiplet 
are experimentally more compressed than the SM predictions, whatever 
the effective interaction, USDA or USDB~\cite{Brow06} . This would be due 
to a weaker residual interaction $\pi d_{5/2} \nu d_{3/2}$ close to the drip line.
It would be interesting to compare these findings 
to calculations taking into account the effect of continuum.  
Note that the present conclusions rely strongly on atomic mass 
determination of the $^{26}_{9}$F ground state, as well as the resonance at
$\sim$270~keV we have assigned to be the $I^\pi=3^+$ state of the multiplet. These
two results deserve to be confirmed.

\section{General conclusions}

The evolution of the $N=28$ shell closure has been investigated 
far from stability, showing a reduction of the spherical shell gap. 
It was found that a progressive onset of 
deformation occurs after the removal of  several protons. Starting
from a rigid-spherical $^{48}_{20}$Ca nucleus,  
$^{46}_{18}$Ar has a vibrational character, $^{44}_{16}$S 
exhibits a shape coexistence,  $^{42}_{14}$Si is likely to 
be oblate and $^{40}_{12}$Mg is expected to be prolate. 
Adding the triaxiality in the $^{48}$Ar nucleus, a wealth of 
nuclear structures can be explored in this mass region.

This structural evolution has been probed using various experimental 
techniques, such as in-beam $\gamma$-ray spectroscopy, transfer, 
Coulomb excitation, g factor measurement; to quote a few. The 
evolution of the $sd$ proton  and $fp$ neutron orbits was probed, from 
which important conclusions were derived. Starting with proton 
orbits, a total disruption of the  $Z=16$ sub-shell, formed 
between the $s_{1/2}$ and $d_{3/2}$ orbits, occurs from $N=20$ to 
$N=28$. In parallel a reduction of the $Z=14$ gap (between the 
$d_{5/2}$ and $s_{1/2}$ orbits) seems to be present as well. 
This global shrink between proton states favors the development 
of pairing and quadrupole excitations at $N=28$ as compared to 
$N=20$, where the magicity persists down to 
$^{34}_{14}$Si. As for neutrons orbits, reductions of the $f$ 
and $p$ SO splittings as well as the $N=28$ shell gap (by about 25\%) 
were proposed  between $^{48}$Ca and $^{42}$Si. This reduction 
is large enough to trigger the development of quadrupole 
excitations across the $N=28$ gap. In all, both the proton  and
 neutron  shell gaps are reduced in the spin-unsaturated $^{42}$Si 
 nucleus, mutually enhancing the collectivity there. 

These studies have in addition shown the many facets of nuclear 
force that can be explored, such as the central, spin-orbit, tensor 
and three-body terms. It also triggered the question of 
proton-neutron forces below $^{42}$Si and more generally when 
approaching the drip-line. These points were addressed in the 
last part of the manuscript. Conclusions are as follows:

\begin{itemize}
\item Due to three-body effects, the $N=28$ gap grows from $N=20$ 
to $N=28$ by about 2.7~MeV. As similar forces seem to be at play 
in the other SO shell gaps, an empirical rule was proposed to 
predict the increase of the $N=14$, $N=50$, $N=82$ shell gaps 
when the $\ell \uparrow$ orbits (with $\ell$ =2,3,4) are completely filled.
Following this rule, a new $N=90$ sub-shell gap should be 
present in $^{140}$Sn, leading to 
interesting consequences for the r-process nucleosynthesis. 

\item The roles of the central, tensor and spin-orbit components 
of the nuclear force were examined in the framework of the shell model 
using various experimental data. The bubble nucleus $^{34}$Si has been used
to probe the reliability of the density and isospin dependence of the SO 
interaction in mean-field models. It was derived that the spin-orbit interaction 
is indeed density dependent, but also seems to have an important isospin 
dependence that is not present in RMF approaches. This feature should
have significant  consequences to model the structure of 
superheavy and neutron-rich nuclei, which exhibit central density 
depletion and surface diffuseness, respectively.

\item Below $^{42}$Si, central forces  further reduce the 
$N=28$ gap. The same situation is expected to occur for the $N=50$ 
and $N=82$ shell gap below $^{78}$Ni and $^{132}$Sn, respectively. 
The reduction of the $N=82$ gap would have extremely important 
consequences for the r-process nucleosynthesis.

\item When reaching the neutron drip-line, proton-neutron forces 
may be weakened by the fact that neutrons are no longer confined 
in the nuclear potential well. The study of $^{26}$F was proposed 
to test such effects. It was found that both the monopole and the residual proton-neutron 
interactions are weakened close to the drip-line as compared to the values used for nuclei close to stability. 
This very preliminary study is foreseen to be a benchmarking case for 
shell-model approaches or coupled cluster models which treat the influence 
of the continuum. 

\end{itemize}

More generally, in the present work we have explored in a semi-quantitative 
manner several facets of the nuclear force which lead to modification of 
shell structure far from stability.  Most of these properties were derived  
from two-body effective interactions (including monopoles) in the shell-model framework.
The \emph{effective} interactions are built from realistic bare interactions 
to which medium effects are added. These interactions are subsequently 
fitted to experimental data to include effects from missing terms such as the three-body components. 
The search for which parts of the nuclear interaction give rise to 
significant shell modifications was started a long time ago 
(e.g. in~\cite{Talmi,Pove81,Fede79}). However very recent works manage 
to point out which properties of the bare nuclear force are preserved 
in the nucleus and which are modified. To quote a few it was  found in 
Refs.~\cite{Otsu10,Smir12} that the tensor force is almost kept constant 
from the bare forces to the nucleus, while the central part of the 
interaction varies a lot. Moreover some theoretical works can now model 
nuclear systems with few nucleons or close to a core nucleus starting 
from realistic interactions or/and treat the interaction with the continuum. 

As for experiments, important results are to come soon in the $N=28$ region,
such as the B(E2) value and the identification of higher-lying 
states in $^{42}$Si, as well as the spectroscopy of neutron-rich 
Mg isotopes when approaching $N=28$. Tremendous progresses have 
been achieved over the last 20 years since the first suggestion 
that the $^{44}$S nucleus was deformed, based on its very short 
lifetime and small neutron delayed emission probability~\cite{Sorl93}, 
to the generalization of nuclear properties
in all SO magic nuclei. All this 
was possible with the increase of experimental capabilities as 
well as regular exchanges between experimentalists and 
theoreticians. 

\subsection{Acknowledgments}
J.-C. Thomas and R. Kanungo are greatly acknowledged for their suggestions and supports as well as for the careful reading of the manuscript. Discussions with F. Nowacki and T. Otsuka were essential for writing the parts of the manuscript related to shell model calculations and monopole decompositions. They are greatly acknowledged for this.

\end{document}